\documentclass[aps,prd,twocolumn,superscriptaddress,nofootinbib,10pt]{revtex4-1}
\usepackage[paperwidth=21cm,paperheight=29.7cm,top=2.54cm,bottom=2.54cm,left=2cm,right=2cm]{geometry}
\usepackage[colorlinks,linkcolor=blue,anchorcolor=blue,citecolor=blue,urlcolor=blue]{hyperref}
\usepackage{graphicx,subfigure}
\usepackage[figuresright]{rotating}
\usepackage{amsmath,amsfonts,amssymb,bm}
\usepackage{array,enumitem,multirow}
\usepackage{acronym}
\newcommand{\be}{\begin{equation}}
\newcommand{\ee}{\end{equation}}
\newcommand{\bea}{\begin{eqnarray}}
\newcommand{\eea}{\end{eqnarray}}
\newcommand{\nn}{\nonumber}

\newcommand{\mSun}{{\rm~M_\odot}}
\newcommand{\TQC}{MOE Key Laboratory of TianQin Mission, TianQin Research Center for Gravitational Physics \&  School of Physics and Astronomy, Frontiers Science Center for TianQin, Gravitational Wave Research Center of CNSA, Sun Yat-sen University (Zhuhai Campus), Zhuhai 519082, China.}

\newacro{GR}{general relativity}
\newacro{GW}{Gravitational wave}
\newacro{MG}{modified graivty theory}
\newacro{PN}{post-Newtonion}
\newacro{ppE}{parameterized post-Einsteinian}
\newacro{GCB}{galactic ultra-compact binary}
\newacro{SBHB}{stellar-mass black hole binary}
\newacro{MBHB}{massive black hole binary}
\newacro{IMBHB}{intermediate-mass black hole binary}
\newacro{EMRI}{extreme mass ratio inspiral}
\newacro{IMRI}{intermediate mass ratio inspiral}
\newacro{SGWB}{stochastic gravitational wave background}
\newacro{CE}{Cosmic Explorer}
\newacro{ET}{Einstein Telescope}
\newacro{LISA}{Laser Interferometer Space Antenna}
\newacro{EdGB}{Einstein-dilaton Gauss-Bonnet}
\newacro{dCS}{dynamic Chern-Simons}
\newacro{SNR}{signal-to-noise ratio}
\newacro{FIM}{Fisher information matrix}
\newacro{ISCO}{innermost stable circular orbit}

\begin{document}
\title{Testing general relativity with TianQin: the prospect of using the inspiral signals of black hole binaries}

\author{Changfu Shi}
\affiliation{\TQC}
\author{Mujie Ji}
\affiliation{\TQC}
\affiliation{Shantou Jinshan Middle School, Shantou 515073, China}
\author{Jian-dong Zhang}
\affiliation{\TQC}
\author{Jianwei Mei}
\email{Email: meijw@sysu.edu.cn (Corresponding author)}
\affiliation{\TQC}

\date{\today}

\begin{abstract}
In this paper, we carry out a systematic study of the prospect of testing general relativity with the inspiral signal of black hole binaries that could be detected with TianQin. The study is based on the parameterized post-Einsteinian (ppE) waveform, so that many modified gravity theories can be covered simultaneously. We consider black hole binaries with total masses ranging from $10\mSun\sim10^7\mSun$ and ppE corrections at post-Newtonian (PN) orders ranging from $-4$PN to $2$PN. Compared to the current ground-based detectors, TianQin can improve the constraints on the ppE phase parameter $\beta$ by orders of magnitude. For example, the improvement at the $-4$PN and $2$PN orders can be about $13$ and $3$ orders of magnitude (compared to the results from GW150914), respectively. Compared to future ground-based detectors, such as ET, TianQin is expected to be superior below the $-1$PN order, and for corrections above the $-0.5$PN order, TianQin is still competitive near the large mass end of the low mass range $[10\mSun\,,\,10^3\mSun]\,$. Compared to the future space-based detector LISA, TianQin can be competitive in the lower mass end as the PN order is increased. For example, at the $-4$PN order, LISA is always superior for sources more massive than about $30\mSun\,$, while at the $2$PN order, TianQin becomes competitive for sources less massive than about $10^4\mSun\,$. We also study the scientific potentials of detector networks involving TianQin, LISA and ET, and discuss the constraints on specific theories such as the dynamic Chern-Simons theory and the Einstein-dilaton Gauss-Bonnet theory.
\end{abstract}

\maketitle

\section{Introduction}

Pushing the experimental limit on testing \ac{GR} is essential in helping find out the breaking point of the century-old theory and revealing the deeper nature of gravity. \ac{GR} has been tested under a variety of conditions, such as with solar system experiments and astrophysical observations \cite{Baker:2014zba}, yet no sure sign of beyond \ac{GR} effect has been found \cite{Will:2014kxa}. \acp{GW} generated from the very early universe or by extremely compact objects such as black holes can help extend the realm of testing \ac{GR} to the genuinely strong field regime. Since the first detection of \acp{GW} by LIGO \cite{LIGOScientific:2016aoc,LIGOScientific:2021djp}, many tests of \ac{GR} have been carried out and all the \ac{GW} data has been found to be consistent with \ac{GR} so far \cite{LIGOScientific:2016lio,LIGOScientific:2019fpa,LIGOScientific:2020tif,LIGOScientific:2021sio,Perkins:2021mhb, Wang:2021jfc,Niu:2021nic, Wang:2021ctl, Kobakhidze:2016cqh,Yunes:2016jcc}.

Space-based \ac{GW} detection in the millihertz frequency band enjoins rich types, large numbers, and diverse spatial distributions of \ac{GW} sources, and expects many \ac{GW} signals that are large in magnitude and/or long in duration \cite{Amaro-Seoane:2012aqc,Baker:2019nia,Hu:2017yoc}. These factors make the millihertz frequency band the golden band in \ac{GW} detection, bearing great importance in fundamental physics \cite{LISA:2022kgy}, astrophysics \cite{Amaro-Seoane:2022rxf} and cosmology \cite{LISACosmologyWorkingGroup:2022jok}. So for a space-based detector, it is of great importance to study its potential in testing \ac{GR} \cite{Gair:2012nm}.

A difficulty in assessing the capability of a space-based \ac{GW} detector in testing \ac{GR} is the lack of a unique direction for the task. The success of \ac{GR} against experimental tests has resulted in a lack of effective guidance in the construction of \acp{MG}, leading to a rather diversified literature that has to be navigated with the help of a mathematical theorem (see, e.g. \cite{Berti:2015itd}). The many different types of \ac{GW} signals expected for a space-based detector only add to the complexity of the task.

There have been a few strategies to deal with the problem. For example, one can focus on testing if the detected \ac{GW} signals are consistent with the predictions of \ac{GR}, such as residual test, inspiral-merge-ringdown coincidence test, polarization test, and so on. One can also employ waveform models that use a set of purely phenomenological parameters to signify possible deviation from \ac{GR} and use the observed data to place constraints on these parameters. Both schemes have been used by the LIGO-Virgo-KAGRA collaboration \cite{LIGOScientific:2016lio,LIGOScientific:2019fpa,LIGOScientific:2020tif,LIGOScientific:2021sio}. For more focused treatment, one can use phenomenological waveforms that are tailored to a chosen set of \acp{MG}, then one not only can place constraints on several \acp{MG} simultaneously, but can also translate the results to an individual \ac{MG} if needed. A good example here is the \ac{ppE} waveform \cite{Yunes:2009ke}, which is based on the \ac{PN} approximation and is most suitable for binary systems in their early inspiral stage and with comparable component masses.

In this paper, we use the \ac{ppE} waveform to carry out a systematic study of the prospect of using TianQin to test \ac{GR}. TianQin is a planned space-based \ac{GW} detector expected around 2035 \cite{Luo:2015,TianQin:2020hid,Tan:2020xbm,Ye:2020tze}. The target frequency band of TianQin is between $10^{-4}\rm$ $\rm Hz$ and 1 $\rm Hz$ \cite{Hu:2018yqb,Zhou:2021psj}, and the expected sources include \ac{GCB} \cite{Huang:2020rjf}, \ac{MBHB} \cite{Wang:2019ryf,Feng:2019wgq}, \ac{IMBHB} \cite{Liu:2021yoy}, \ac{EMRI} \cite{Fan:2020zhy}, \ac{SBHB} \cite{Liu:2020eko}, and \ac{SGWB} \cite{Liang:2021bde}. There might also be unexpected sources \cite{TianQin:2020hid,Fan:2022wio}. A series of work has been carried out to assess the scientific potential of TianQin, such as on studying the astrophysical history of galaxies and black holes \cite{Huang:2020rjf,Wang:2019ryf}, the dynamics of dense star clusters and galactic centers \cite{Fan:2020zhy}, the nature of gravity and black holes \cite{Shi:2019hqa,Bao:2019kgt,Zi:2021pdp,Sun:2022pvh,Xie:2022wkx}, the expansion history of the universe \cite{Zhu:2021aat,Zhu:2021bpp}, and the fundamental physics related to the very early universe \cite{Wang:2021dwl,Wang:2020jrd,Wei:2022poh}. This work is part of the effort.

Apart from doing a broad test of \ac{GR} by using the \ac{ppE} waveform, we study how the results look like for individual \acp{MG}. For this purpose we use two theories as examples: the \ac{dCS} theory and the \ac{EdGB} theory. There is no particular reason why these two theories are chosen, apart from the fact that the \ac{ppE} waveforms are known in these theories.

We also carry out a parallel study of some other detectors as a comparison and to figure out the scientific potential of detector networks made of these detectors. Important examples include the third generation ground-based detectors, \ac{CE} \cite{Reitze:2019iox} and \ac{ET} \cite{Punturo:2010zz}, and the space-based detector, LISA \cite{Audley:2017drz}. Since there have been results on the joint detection of TianQin and \ac{CE} \cite{Carson:2019rda}, we focus on \ac{ET} and LISA in this paper.

The paper is organised as following. In section \ref{sec:existing.work}, we summarise the main existing works that are related to this one. In section \ref{sec:ppe}, we recall the basic results on the \ac{ppE} waveform. In Section \ref{sec:method}, we present the methods and key assumptions used in the calculations. In sections \ref{sec:rst-ppE} and \ref{sec:resultsII}, we present our main findings. The paper concludes with a summary in section \ref{sec:summary}. Throughout this paper, we use the natural units in which $G_N=\hbar=c=1\,$.

\section{Summary of existing results}\label{sec:existing.work}

A lot of works have already been done on using the inspiral signals detected by the space-based detector LISA to test \ac{GR}. Early works included using signals from extreme mass ratio inspiral systems to test the no-hair theorem \cite{Ryan:1995wh,Ryan:1997hg} and using signals from neutron stars inspiraling into intermediate-mass black holes to test the Scalar-Tensor theory \cite{Scharre:2001hn}.

For systems with comparable component masses, Berti et al. have considered using inspiral signals to constrain the massive Brans-Dicke theory by introducing leading order corrections to the \ac{PN} waveform \cite{Berti:2004bd}. Arun et al. have used a set of phenomenological phase parameters (one for each \ac{PN} order) to characterize the deviation of an \ac{MG} from \ac{GR} \cite{Arun:2006yw} and placed constraints on these phenomenological phase parameters \cite{Arun:2006yw,Arun:2006hn}. This is the precursor to the \ac{ppE} method \cite{Yunes:2009ke}, which uses a new set of phenomenological parameters to replace the phenomenological phase parameters, by dividing out the corresponding velocity factor at each \ac{PN} order.

Connish et al. have studied how the \ac{ppE} parameters can be constrained by future detectors, such as aLIGO/aVirgo and LISA \cite{Cornish:2011ys}. Huwyler et al. have investigated the potential of using LISA to constrain the ppE phase parameter $\beta$, as to be defined in (\ref{eq:waveform}), with \ac{MBHB} \cite{Huwyler:2014vva}. The \ac{ppE} formalism has also been used to place constraints on specific \acp{MG}, such as Brans-Dicke theory \cite{Zhang:2017sym}, Lorentz-Violating Gravity \cite{Hansen:2014ewa}, G(t) theory \cite{Yunes:2009bv}, and theories with massive gravitons, modified dispersion relations or dipole radiation \cite{Keppel:2010qu,Mirshekari:2011yq,Berti:2011jz,Samajdar:2017mka,Arun:2012hf}.

After the direct detection of \acp{GW}, Yunes et al. have analyzed the constraints on the ppE phase parameters using the GW190514 and GW151226 signals, and have translated the results to some specific \acp{MG} \cite{Yunes:2016jcc}. Chamberlain et al. have studied how some future detectors (four possible configurations of LISA, aLIGO, A+, Voyager, CE, and ET-D) can constrain the ppE phase parameter $\beta$ and some \acp{MG} (including dipole radiation, extra dimensions, G(t) theory, Einstein-\AE ther theory, Khronometric gravity and Massive graviton theory), by using some example \ac{GW} signals \cite{Chamberlain:2017fjl}.

After the multiband work on \ac{SBHB} by Sesana \cite{Sesana:2016ljz}, Barausse et al. have employed the \ac{ppE} formalism to show that the multiband observation with aLIGO and LISA can improve the expected constraints on the \ac{GW} dipole radiation by 6 orders of magnitude \cite{Barausse:2016eii}, Carson et al. have studied constraints on the \ac{ppE} parameters with multiband observation using \ac{CE} and several space-based detectors (LISA, TianQin, DECIGO and B- DECIGO) \cite{Carson:2019rda}, and they have also analyzed the multiband enhancement on constraining the \ac{EdGB} theory and the IMR consistency test \cite{Carson:2020cqb}.

Comparing to these existing works, we will do a more thorough exploration on how the constraints on the ppE parameters will depend on different source parameters, different detectors, different detection schemes, and possibly, also different detector networks. 

\section{The parameterized post-Einsteinian waveform}\label{sec:ppe}

Black holes binaries are ideal systems for testing \ac{GR}, for the strong field condition they can provide and for the less of environmental contamination that often affects other astrophysical systems. The evolution of a black hole binary can be divided into three phases: inspiral, merger, and ringdown. During the inspiral phase, the two components of the system start widely separated and their velocities are relatively small. The corresponding waveforms can be well modeled through the \ac{PN} approximation for systems with comparable component masses. In \ac{GR}, the frequency domain waveform is given by
\bea&&h_{\rm GR}(f)=A(f)e^{i\psi(f)}\notag\,,\\
&&\psi(f)=2\pi t_c +\phi_c+\Sigma_{k=0}^\infty\phi_k^{\rm PN} u^{(k-5)/3}\,,\label{pna}\eea
where $f$ is frequency, $A(f)$ is the amplitude, $t_c$ and $\phi_c$ are the coalescence time and phase, respectively, $u=(\pi\mathcal{M}f)^{1/3}$ is a characteristic velocity, $\mathcal{M}=\eta^{3/5}M$ is the chirp mass,  $M=m_1+m_2$ is the total mass, $\eta=m_1m_2/(m_1+m_2)^2$ is the symmetrical mass ratio, and $\phi^{\rm PN}_k$ is the phase coefficient at the $(k/2)$PN order. Note $\phi^{\rm PN}_k$ is completely determined by the source parameters for a binary black hole system \cite{Blanchet:2002av}.

The \ac{ppE} waveform has been proposed \cite{Yunes:2009ke} to study \acp{MG} whose inspiral waveform has the same \ac{PN} structure as (\ref{pna}). The difference between a given \ac{MG} and \ac{GR} resides in how the amplitude and the coefficients $\phi^{\rm PN}_k$ depend on the source parameters. Suppose the \ac{MG} correction only happens at a particular \ac{PN} order or keeping only the leading order correction, the waveform is given by
\be h_{\rm ppE}(f)=h_{\rm GR}(f)(1+\alpha u^a)e^{i\beta u^b}\,,\label{eq:waveform}\ee
where $\alpha$ and $\beta$ are the \ac{ppE} parameters, and $a$ and $b$ are the \ac{ppE} order parameters, satisfying
\bea b=2{\rm~ PN}-5\,,\quad a=b+5\,.\eea
\ac{GR} is recovered with $\alpha=\beta=0\,$.

The original work of \cite{Yunes:2009ke} has only considered the two \ac{GW} polarization modes found in \ac{GR} and has focused on quasi-circular orbits for the black hole binaries. Extensions have been made to include extra polarization modes \cite{Chatziioannou:2012rf}, time domain waveforms \cite{Huwyler:2014gaa}, eccentricity \cite{Loutrel:2014vja} and environmental effect \cite{Cardoso:2019rou}. For any particular \ac{MG}, the relation between the theory and the \ac{ppE} parameter can be established by calculating corrections to the evolution of the binary orbits \cite{Tahura:2018zuq}. In this way, the \ac{ppE} parameters have been calculated for a series of theories, such as Brans-Dicke gravity \cite{Zhang:2017sym}, screened modified gravity \cite{Zhang:2017srh}, parity-violating gravity \cite{Zhao:2019xmm}, Lorentz-violating gravity \cite{PhysRevD.91.082003}, noncommutative gravity \cite{PhysRevD.94.064033}, and quadratic modified gravity \cite{Yagi:2011xp}. For the \ac{EdGB} and \ac{dCS} theories that will be considered in this paper, the \ac{ppE} parameters have also been calculated \cite{Yagi:2011xp}.

The leading order modification from \ac{EdGB} starts at the $-1$PN order, corresponding to $b=-7$ and $a=-2$. The \ac{ppE} parameters are \cite{Yagi:2011xp}:
\begin{align}
\alpha_{\rm EdGB}&=-\frac{5\zeta_{\rm EdGB}}{192}\frac{(m_1^2\tilde{s}_2-m_2^2\tilde{s}_1)^2}{M^4\eta^{18/5}}\,,\nn\\
\beta_{\rm EdGB}&=-\frac{5\zeta_{\rm EdGB}}{7168}\frac{(m_1^2\tilde{s}_2-m_2^2\tilde{s}_1)^2}{M^4\eta^{18/5}}\,,
\label{EdGB}
\end{align}
where $\zeta_{\rm EdGB}\equiv16\pi\bar{\alpha}^2_{\rm EdGB}/M^4\,$, $\bar{\alpha}_{\rm EdGB}$ is the coupling between the scalar field and quadratic curvature term in the theory \cite{Kanti:1995vq}, and $\tilde{s}_n\equiv2(\sqrt{1-\chi_n^2}-1+\chi_n^2)/\chi_n^2\,$, $n=1,2$, is the spin-dependent scalar charge of the $n$-th component, with $\chi_n$ being the effective spin. The current best constraint on the theory comes from the observation of GW200115, giving $\sqrt{|\bar{\alpha}_{\rm EdGB}|}<1.3$ km \cite{Lyu:2022gdr}.

The leading order modification from \ac{dCS}  starts at the 2PN order, corresponding to $b=-1$ and $a=4$. The \ac{ppE} parameters are \cite{Yagi:2011xp,Tahura:2018zuq}:
\bea \alpha_{\rm dCS}&=\frac{57713\eta^{-14/5}\xi_{\rm dCS}}{344064}\Big[\Big(1-\frac{14976\eta}{57713}\Big)\chi_a^2\nn\\
&+\Big(1-\frac{215876\eta}{57713}\Big)\chi_s^2-2\delta_m\chi_a\chi_s\Big]\,,\nn\\
\beta_{\rm dCS}&=-\frac{1549225\eta^{-14/5}\xi_{\rm dCS}}{11812864}\Big[\Big(1-\frac{16068\eta}{61969}\Big)\chi_a^2\nn\\
&+\Big(1-\frac{231808\eta}{61969}\Big)\chi_s^2-2\delta_m\chi_a\chi_s\Big]\,,\label{dCS}
\eea
where  $\delta_m\equiv(m_1-m_2)/M\,$, $\chi_s=(\chi_1+\chi_2)/2\,$, $\chi_a=(\chi_1-\chi_2)/2\,$, $\xi_{\rm dCS}\equiv16\pi\bar{\alpha}^2_{\rm dCS}/M^4$, and $\bar{\alpha}_{\rm dCS}$ is the coupling constant of the Chern-Simons correction \cite{Jackiw:2003pm}. The current best constraint on the theory comes from the observation of neutron star systems, giving $\sqrt{\bar{\alpha}_{\rm dCS}}<8.5$ km \cite{Silva:2020acr}. So far one is unable to place a meaningful constraint on the \ac{dCS} theory using \ac{GW} data directly, due to a lack of viable waveform.

\section{Methods and assumptions}\label{sec:method}

We use the \ac{FIM} method to estimate the constraints on the \ac{ppE} parameters $\alpha$ and $\beta$, and on the theory specific couplings $\bar{\alpha}_{\rm EdGB}$ and $\bar{\alpha}_{\rm dCS}\,$. The whole parameter space is given by
\bea \vec{\theta}=\{M,\eta,D_L,t_c,\phi_c,\chi_1,\chi_2,\theta_{nonGR}\}\,,\label{parameter}\eea
where $D_L$ is the luminosity distance, and $\theta_{nonGR}$ stands for the non-GR parameters such as $\alpha\,$, $\beta\,$, $\bar\alpha_{\rm EdGB}$ and $\bar\alpha_{\rm dCS}\,$.

Assuming large \ac{SNR} and Gaussian noise, the uncertainties in the waveform parameters are characterized by
\bea \Delta\theta^a\equiv\sqrt{\langle \Delta\theta^a\Delta\theta^a\rangle}\approx\sqrt{(\Gamma^{-1})^{aa}}\,,\label{estimation}\eea
where $\langle\dots\rangle$ stands for statistical average and $\Gamma^{-1}$, the covariance matrix, is the inverses of \ac{FIM} \cite{Finn:1992,Cutler:1994},
\bea\Gamma_{ab}=\Big(\frac{\partial h}{\partial \theta^a}\Big|\frac{\partial h}{\partial \theta^b}\Big)\,.\label{FIM}\eea
When a signal is observed by multiple detectors simultaneously, the combined \ac{FIM} is
\bea\Gamma_{ab}^{total}=\Gamma_{ab}^{(1)}+\Gamma_{ab}^{(2)}+\dots\,,\eea
where $1,2,\dots$ denote different detectors.

The inner product in (\ref{FIM}) is defined as
\bea(p|q)\equiv2\int_{f_{\rm low}}^{f_{\rm high}} \frac{p^*(f)q(f)+p(f) q^*(f)}{S_n(f)}df\,,\label{inner product}\eea
where $S_n(f)$ is the sensitivity of the detector. The low- and high-frequency cutoffs are taken to be:
\bea f_{\rm low}&=& \max\Big[f_{\rm low}^{\rm PN}\,,\,f_{\rm low}^D\Big]\,,\nn\\
f_{\rm high}&=&\min\Big[f_{\rm ISCO}\,,\,f_{\rm high}^D\Big]\,,\eea
where $f_{\rm low}^D$ and $f_{\rm high}^D$ mark the end points of the sensitivity band of the detector, $f_{\rm ISCO}=(6^{3/2}\pi M)^{-1}$ is the frequency at the \ac{ISCO}, and
\bea f_{\rm low}^{\rm PN}=(8\pi \eta^{3/5}M)^{-1}(5\eta^{3/5}M/T_{ob})^{3/8}\,
\label{eq:fbandpn}
\eea
is determined by the total observation time $T_{ob}\,$. In this paper, we will take $T_{ob}$ to be the length of time from the beginning of the observation to the moment when the binaries reach \ac{ISCO}.

For the detectors we consider TianQin \cite{Hu:2018yqb}, LISA \cite{Robson:2018ifk}, ET \cite{Punturo:2010zz}, and the twin constellation configuration of TianQin \cite{Shi:2019hqa,Wang:2019ryf}.
The sky averaged Michelson sensitivity of TianQin can be modeled as \cite{Hu:2018yqb,Wang:2019ryf},
\bea S_n(f)=\frac{10}{3}\Big[1+\Big(\frac{2fL_0}{0.41}\Big)^2\Big]S_N(f)\,,\label{curveTianQin}\eea
where we use the following noise model \cite{Hu:2018yqb,Huang:2020rjf,Liang:2021bde}:
\bea S_N(f)=\frac1{L_0^2}\Big[\frac{4S_a}{(2\pi f)^4}\Big(1+\frac{10^{-4}{\rm Hz}}{f}\Big)+S_x\Big]\,.\label{noiseTianQin}\eea
Here $L_0=\sqrt{3}\times 10^8~{\rm m}$ is the arm length, $\sqrt{S_a}=1\times 10^{-15}{\rm~m/s^2/Hz^{1/2}}$ is the residual acceleration on each test mass, and $\sqrt{S_x}=1\times10^{-12}~{\rm m/Hz^{1/2}}$ is the displacement measurement noise in each laser link. The sensitivity band of the detectors are chosen as
\bea f_{\rm low}^D&=&10^{-4}~{\rm Hz}\,,\quad f_{\rm high}^D=1~\rm Hz\,,\;\text{for TianQin}\,,\nn\\
f_{\rm low}^D&=&10^{-6}~{\rm Hz}\,,\quad f_{\rm high}^D=1~\rm Hz\,,\;\text{for LISA}\,,\nn\\
f_{\rm low}^D&=&1~{\rm Hz}\,,\quad f_{\rm high}^D=10^4~\rm Hz\,,\;\text{for ET}\,.\label{eq:fbandd}\eea

All detectors are limited to one year of operation, except in part of subsection \ref{bestsource},  when the effect of $T_{ob}$ is discussed. For TianQin, all binaries used in the calculation are assumed to reach their \acp{ISCO} right when TianQin finishes a 3 month observation (except in subsection \ref{33scheme}, which is dedicated to cases when \acp{ISCO} are reached when TianQin is in between observation time windows). So only the last $0\sim3$ months and $6\sim9$ months data will be used for TianQin. The frequency bounds in the integrals are modified accordingly.

The \ac{GR} waveform $h_{\rm GR}$ in (\ref{eq:waveform}) is generated using IMRPhenomD \cite{Khan:2015jqa,Husa:2015iqa}. We take $t_c=0$, $\phi_c=0$, $\chi_1=0.4$ and $\chi_2=0.2$ in all the calculations, and we only consider \ac{ppE} corrections starting from the \ac{PN} orders in $\{-4$PN, $-3.5$PN, $\cdots\,$, $2$PN\}, corresponding to $b\in[-13,-1]\,$, and black hole binaries with total masses in the range $[10\mSun\,,\,10^7\mSun]\,$. Only sources in the lower mass end will be observable by \ac{ET}, so we roughly divide the mass range into two sectors: the low mass range, $M \in[10\mSun\,$, $10^3\mSun]\,$, and the high mass range, $M \in[10^3\mSun\,$, $10^7\mSun]\,$. All plots in this paper will be made separately for these two mass ranges.

A laser interferometer type detector is more sensitive to the phase of a \ac{GW} signal, so the \ac{ppE} parameter $\beta$ is more severely constrained, while only in limited cases that the effect of the parameter $\alpha$ is not negligible \cite{Tahura:2019dgr}. So for most part of the discussion, we will focus on the constraints on $\beta\,$, while only in subsection \ref{sec:amp} that we will discuss the effect of the amplitude correction parameter $\alpha\,$.

\section{Projected constraints on $\beta$}\label{sec:rst-ppE}

In this section, we discuss the expected constraints on the \ac{ppE} parameter $\beta$. Our main findings are the following.

\subsection{What kind of sources are the best for constraining $\beta$?}
\label{bestsource}

Although all components of the \ac{FIM} contribute to Eq.(\ref{estimation}), the dominant contribution comes from
\bea \Gamma^{(b)}_{\beta\beta}&=&\Big(\frac{\partial h}{\partial \beta}\Big|\frac{\partial h}{\partial \beta}\Big)
=4\int_{f_{\rm low}}^{f_{\rm high}}\frac{u^{2b}h_{\rm GR}h^*_{\rm GR}}{S_n(f)}df\nn\\
&\simeq&\frac{5\pi^{2b-4/3}}{24 D_L^2}\eta^{(5+2b)/5}M^{(5+2b)/3}\nn\\
&&\times\int_{f_{\rm low}}^{f_{\rm high}}\frac{f^{(2b-7)/3}}{S_n(f)}df\,.\label{eq:beta}\eea
One can see that $M$, $\eta$, $D_L$ and $T_{ob}$ are the main parameters affecting the constraints on $\beta\,$, and the effect may differ for different \ac{PN} orders. The luminosity distance $D_L$ contributes rather trivially through an overall scaling,
\bea\Delta\beta=\sqrt{(\Gamma^{-1})_{\beta\beta}}\varpropto D_L\,,\label{eq:dldp}\eea
and so we will not consider it any longer.

The total mass contributes to $\Delta\beta$ through two places. One is through the factor,
\bea\Delta\beta^{(b)}\approx\sqrt{(\Gamma_{(b)}^{-1})_{\beta\beta}}\sim M^{-(5+2b)/6}\,,\label{masssdppl}\eea
which improves with $M$ monotonically for $b=-2,-1$ and worsens with $M$ monotonically for $b<-2\,$.\footnote{Here ``improve" means that the value of $\Delta\beta^{(b)}$ is decreasing, and ``worsen" means that the value of $\Delta\beta^{(b)}$ is increasing.} The other is through the bounds in the integral,
\bea \Delta\beta^{(b)}\sim\bigg(\int_{f_{low}}^{f_{high}}\frac{f^{(2b-7)/3}df}{S_n(f)}\bigg)^{-1/2}\,,\label{massdpint}\eea
which does not have a clear trend and is different for different detectors.

\begin{figure}
\centering
\includegraphics[width=0.48\textwidth]{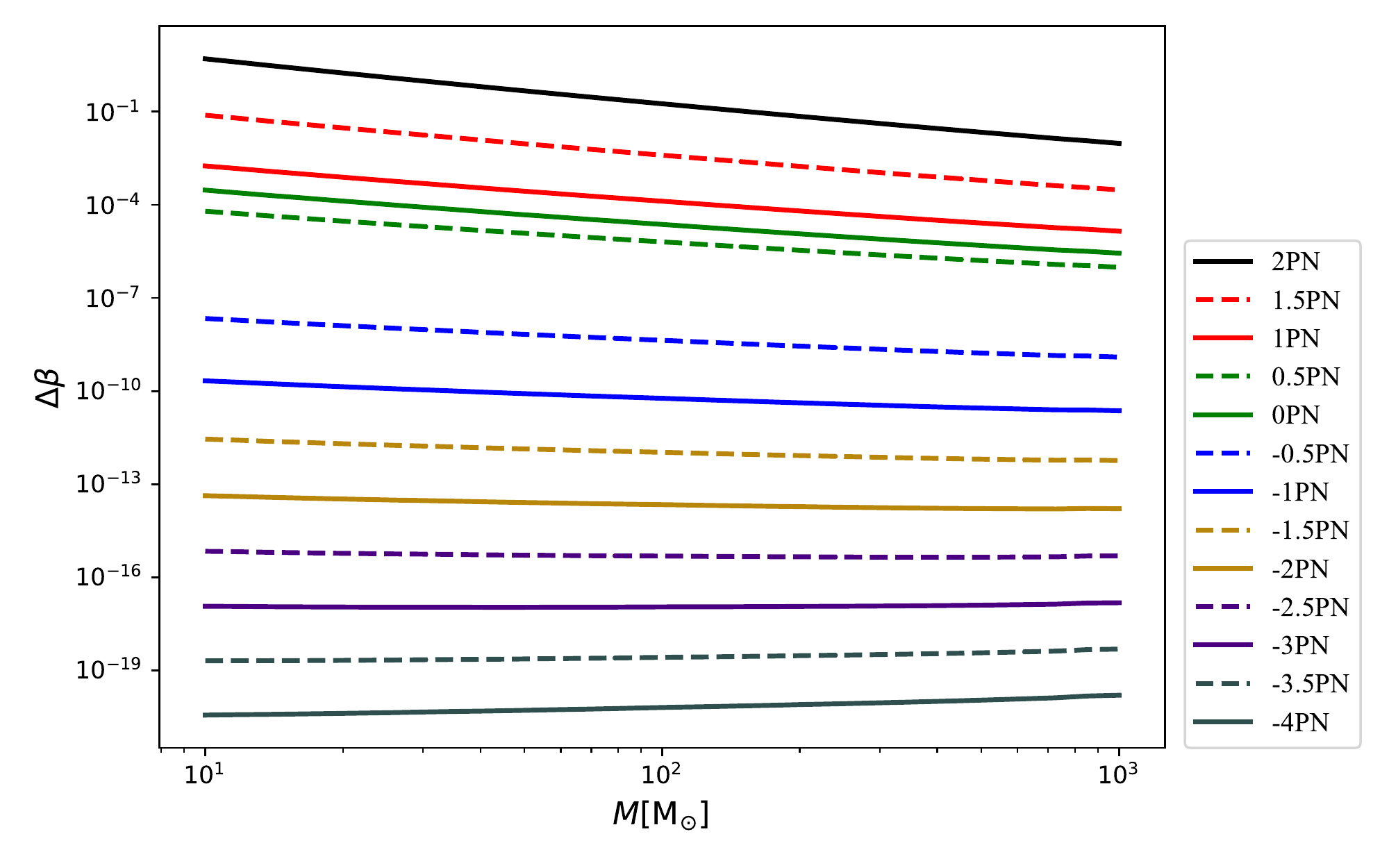}
\caption{Dependence of $\Delta\beta$ on $M$ at different PN order for TianQin, with sources in the low mass range.}
\label{tqstmassdp}
\end{figure}

\begin{figure}
\centering
\includegraphics[width=0.48\textwidth]{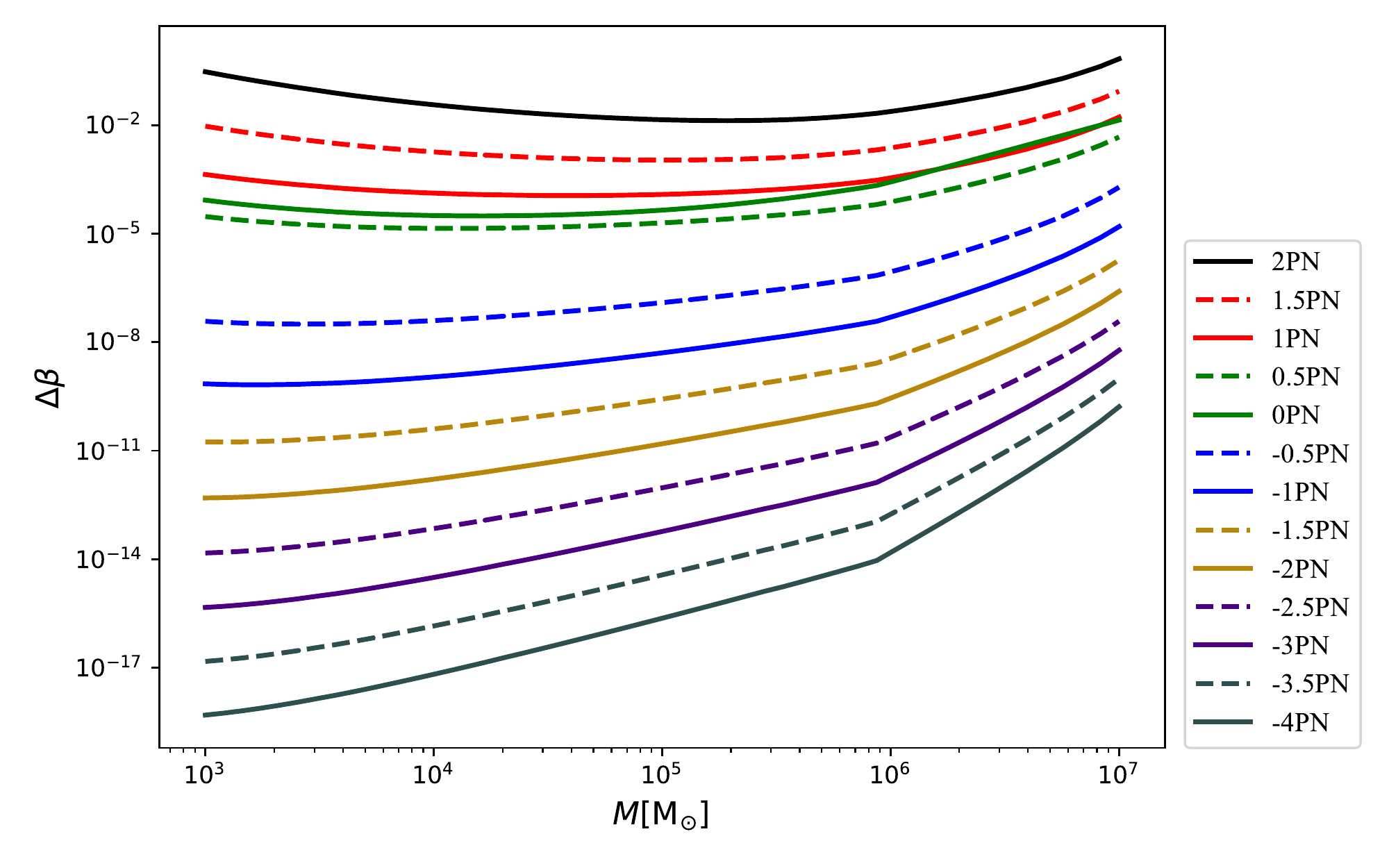}
\caption{Dependence of $\Delta\beta$ on $M$ at different PN order for TianQin, with sources in the high mass range.}
\label{tqspmassdp}
\end{figure}

The dependences of $\Delta\beta$ on $M$ at different PN orders are shown in FIG. \ref{tqstmassdp} and  FIG. \ref{tqspmassdp}. For all the plots, we take $D_L=500$ $\rm Mpc$, $\eta=0.22$ for sources in the low mass range, and $D_L=15$ $\rm Gpc$, $\eta=0.22$ for sources in the high mass range.

One can see that the impact of the total mass on the constraint can be significant. For example, at the 2PN order, the difference in the low mass range can reach three orders of magnitude, while for the $-4$PN order, the difference in the high mass range can reach more than eight orders of magnitude. One can also see that $\Delta\beta$ in the low PN order case is better constrained with low mass sources while that in the high PN order case is better constrained with high mass sources.

One can conclude from \figurename{ \ref{tqstmassdp}} and \figurename{ \ref{tqspmassdp}}  that the ratios of $\Delta\beta$ between adjacent orders with same source tend to be roughly a constant.  Using \eqref{eq:beta}, such ratio can be given:
\bea
\label{ratio}
\frac{\Delta\beta^{(b)}}{\Delta\beta^{(b+1)}}
\approx\sqrt{\frac{\int_{f_{\rm low}}^{f_{\rm high}} u^2\frac{f^{(2b-7)/3}}{S_n(f)}df}{\int_{f_{\rm low}}^{f_{\rm high}}\frac{f^{(2b-7)/3}}{S_n(f)}df}}\approx \bar{u},
\eea
where $\bar{u}$ can be regarded as the weighted-average velocity during the whole process with frequency $f\in [
f_{\rm low}^{\rm PN}, f_{\rm ISCO}]$, corresponding to
\bea
\label{udp}
u\in [\frac{1}{8^{1/3}}(\frac{5\eta^{3/5}M}{T})^{1/8},\frac{\eta^{1/5}}{6^{1/2}}].
\eea
$u$ are roughly in the order of $\mathcal{O}(10^{-2})$ to $\mathcal{O}(10^{-1})$. There exist an abnormal situation in 0PN  which is caused by a strong correlation with mass, such phenomenon is also mentioned in \cite{Chamberlain:2017fjl}.

\begin{figure}
\centering
\includegraphics[width=0.48\textwidth]{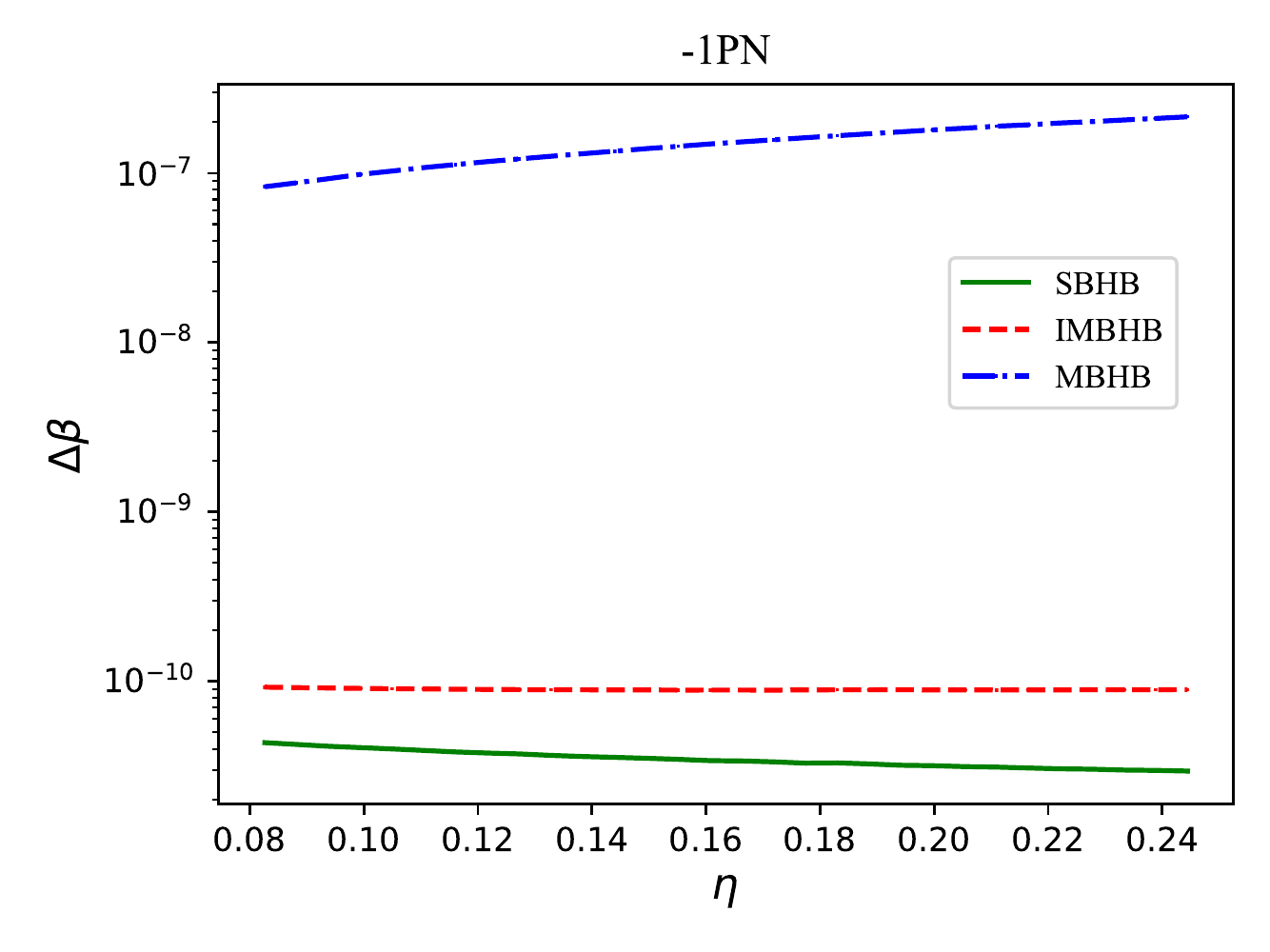}
\caption{Dependence of $\Delta\beta$ on $\eta$ at the $-1$PN order for TianQin.}
\label{mrdp}
\end{figure}

The symmetric mass ratio $\eta$ also contributes to $\Delta\beta$ through two places. One is through the factor
\bea\Delta\beta^{(b)}\approx\sqrt{(\Gamma_{(b)}^{-1})_{\beta\beta}}\propto \eta^{-(5+2b)/10}\,,\label{etadppl}\eea
which improves with $\eta$ monotonically for $b=-2,-1$ and worsens with $\eta$ monotonically for $b<-2\,$. The other is through the low-frequency cutoff,
\bea f_{\rm low}^{\rm PN}\propto\eta^{-3/8}\,,\label{etaflc}\eea
which improves the constraints monotonically with growing $\eta\,$.

The dependences of $\Delta\beta$ on $\eta$ at different PN orders are shown in FIG. \ref{mrdp} and in FIG. \ref{fig.etadpall}. Three sources have been used as examples:
\begin{itemize}
\item \ac{SBHB}: $M=70{\mSun}\,$, $D_L=200$ $\rm Mpc\,$;
\item \ac{IMBHB}: $M=2\times 10^3{\mSun}\,$, $D_L=2$ Gpc;
\item \ac{MBHB}: $M=2\times10^6{\mSun}\,$, $D_L=15$ Gpc.
\end{itemize}
One can see that the constraints vary mildly with $\eta\,$. One can also see that, for \ac{MBHB} and \ac{SBHB}, the variation is mostly dominated by (\ref{etadppl}) and (\ref{etaflc}), respectively.

\begin{figure}
\centering
\includegraphics[width=0.48\textwidth]{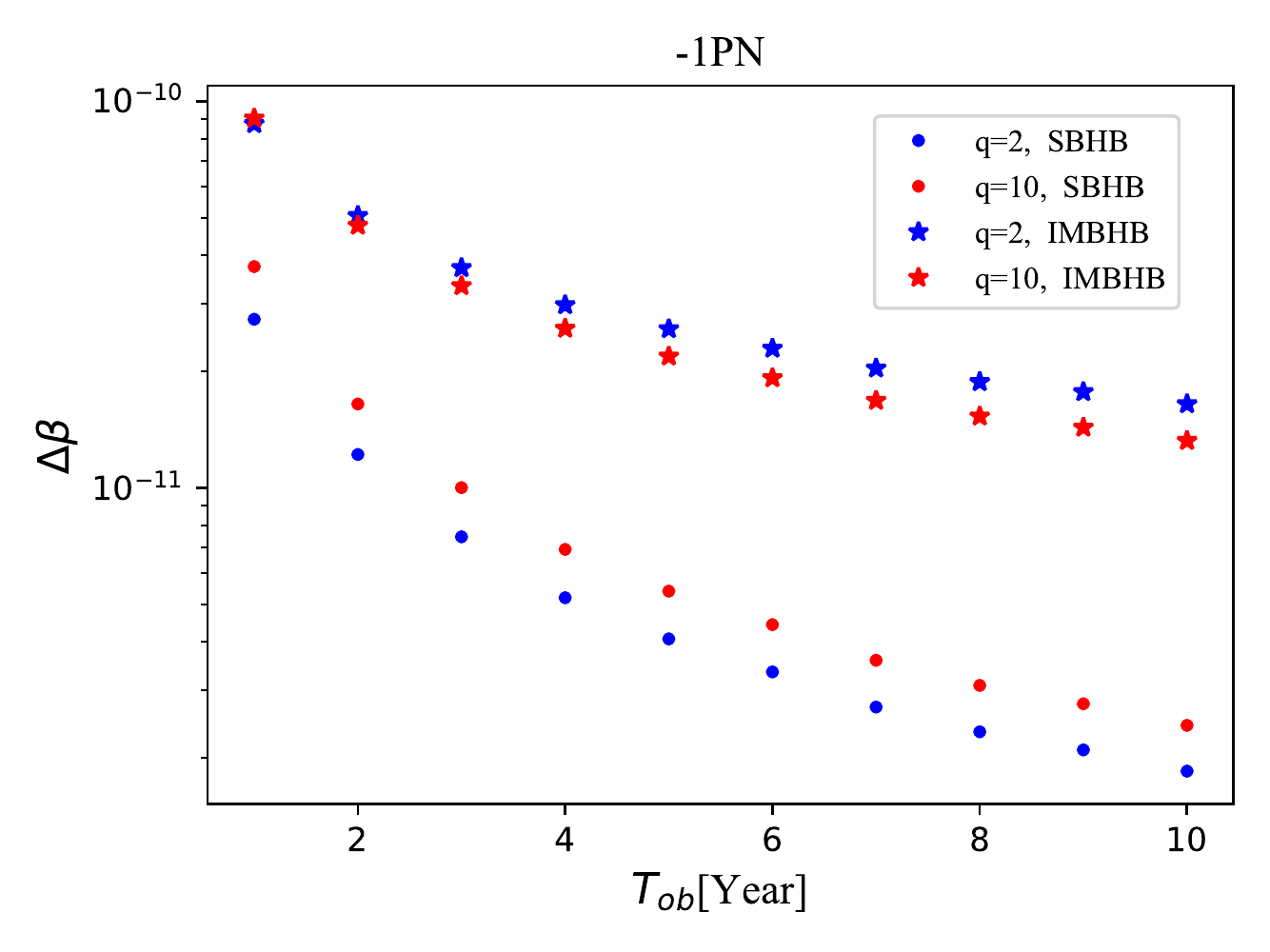}
\caption{Dependence of $\Delta\beta$ on $T_{ob}$ at the $-1$ PN order for TianQin.}
\label{tobdp}
\end{figure}

The total observation time $T_{ob}$ contributes to $\Delta\beta$ through the low-frequency cutoff,
\bea f_{\rm low}^{\rm PN}\propto T_{ob}^{-3/8}\,.\label{eq:tdp}\eea
The dependences of $\Delta\beta$ on $T_{ob}$ at different PN orders are shown in FIG. \ref{tobdp} and in FIG. \ref{fig.tobdpall}. Four sources have been used as examples:
\begin{itemize}
\item \ac{SBHB}: $M=70{\mSun}\,$, $D_L=200$ Mpc, with $q=2$ or $q=10\,$;
\item \ac{IMBHB}: $M=2\times10^3{\mSun}\,$, $D_L=2$ Gpc, with $q=2$ or $q=10\,$.
\end{itemize}
Here $q=m_1/m_2$ is the mass ratio. Ten different values of $T_{ob}$ have been considered, ranging from six months to five years.

One can see that $T_{ob}$ can have significant impact on $\Delta\beta\,$. The effect is particularly strong at lower PN orders with low mass sources. For example, at the -4PN order, increasing the observation time from six month to five years can improve the constraint from a low mass source by about two orders of magnitude.

To summarize, not considering the true abundance of sources at different astrophysical distances, the best sources to constrain $\beta$ at the positive PN orders is \ac{MBHB}, with masses $M>10^4 $$\mSun$; the best sources to constrain $\beta$ at PN orders $[-3$PN, $-0.5$PN] is \ac{IMBHB}, with masses $M\in [10^2\mSun\,$, $10^4\mSun]$; and the best sources to constrain $\beta$ at PN orders below $-3$PN is \ac{SBHB}, with masses $M<10^2\mSun$. In comparison to the results of GW150914 \cite{Yunes:2016jcc}, TianQin would improve the constraints by many orders of magnitude, ranging from $13$ orders of magnitude at the $-4$PN order to about 3 orders of magnitude at the $2$PN order.

\subsection{How does the special detection scheme of TianQin affect the constraints on $\beta$?}\label{33scheme}

The basic concept of TianQin envisions a ``3 months on + 3 months off" observation scheme, in order to cope with the problem of varying solar angles on the spacecraft \cite{Luo:2015}.
Some of the black hole binaries may merge when TianQin is transiting from one observation time window to the next, resulting in a loss of information.

For black hole binaries merging during the non-observation period, the integration bounds of \eqref{eq:beta} need to be modified,
\begin{align}
\label{Tm:fhigh}
f_{\rm high}^{\rm PN}=(8\pi \eta^{3/5}M)^{-1}(5\eta^{3/5}M/T_m)^{3/8}\
\end{align}
and
\begin{align}
\label{Tm:flow}
f_{\rm low}^{\rm PN}=(8\pi \eta^{3/5}M)^{-1}(5\eta^{3/5}M/(T_0+T_m))^{3/8},\
\end{align}
where we take $T_0=9$ month and $T_m$ is the length of missed observation time.

\begin{figure}
\centering
\includegraphics[width=0.48\textwidth]{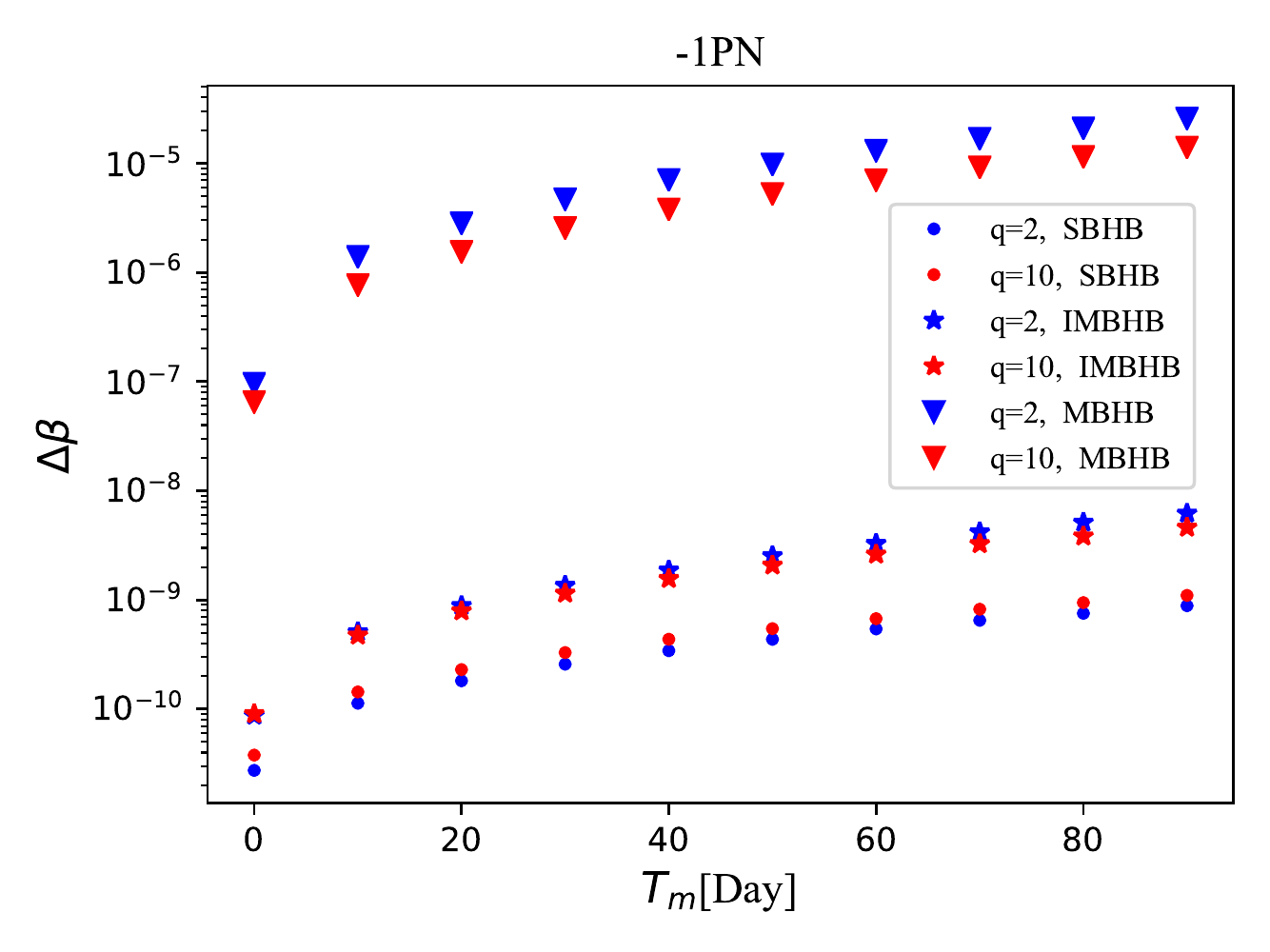}
\caption{Dependence of $\Delta\beta$ on $T_m$ at the $-1\rm PN$ order for TianQin.}
\label{tmdp}
\end{figure}

The dependences of $\Delta\beta$ on $T_m$ at different PN orders are shown in FIG. \ref{tmdp} and in FIG. \ref{fig.tmdpall}. Six sources have been used as examples:
\begin{itemize}
	\item \ac{SBHB}: $M=70{\mSun}\,$, $D_L=200$ Mpc, with $q=2$ or $q=10\,$;
	\item \ac{IMBHB}: $M=2\times10^3{\mSun}\,$, $D_L=2$ Gpc, with $q=2$ or $q=10\,$;
	\item \ac{MBHB}: $M=2\times10^6{\mSun}\,$, $D_L=15$ Gpc, with $q=2$ or $q=10\,$.	
\end{itemize}
Ten different values of $T_m$ have been considered, ranging from zero to three months.

One can see that the effect of $T_m$ is more significant at higher PN orders and for more massive sources. The worst case scenario is when all the last three months of data right before \ac{ISCO} is lost. For \ac{MBHB}, this would mean a lost of more than $99\%$ of the \ac{SNR} \cite{Wang:2019ryf}, making the signal hardly detectable. If the signal is still detectable, then the constraints on $\beta$ would be worsen by about three orders of magnitude. For \ac{IMBHB} and \ac{SBHB}, there will be some amount of \ac{SNR} left for the signals (for example, about 40\% and 76\% \acp{SNR} will be left for the above mentioned \ac{SBHB} and \ac{IMBHB}  sources with $q=2$, respectively), while the constraints on $\beta$ will be worsened by about $1\sim2$ orders at the negative PN orders.

\subsection{How much can a detector network improve on the constraints on $\beta$?}

\begin{figure}
\centering
\includegraphics[width=0.48\textwidth]{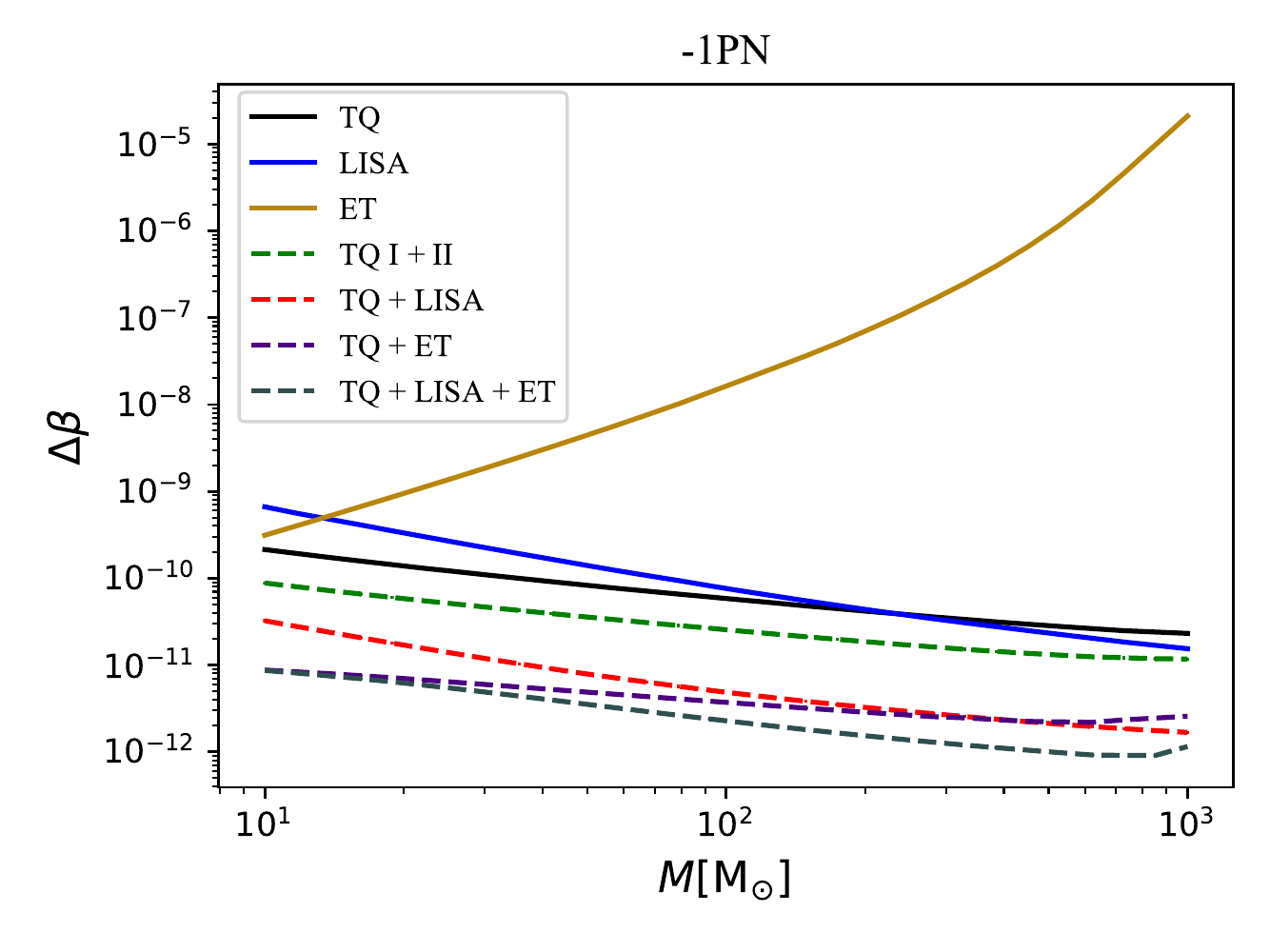}
\caption{Dependence of $\Delta\beta$ on $M$ at the $-1$ PN order for different detector configurations, with sources in the low mass range.}
\label{stmassdp-1}
\end{figure}

\begin{figure}
\centering
\includegraphics[width=0.48\textwidth]{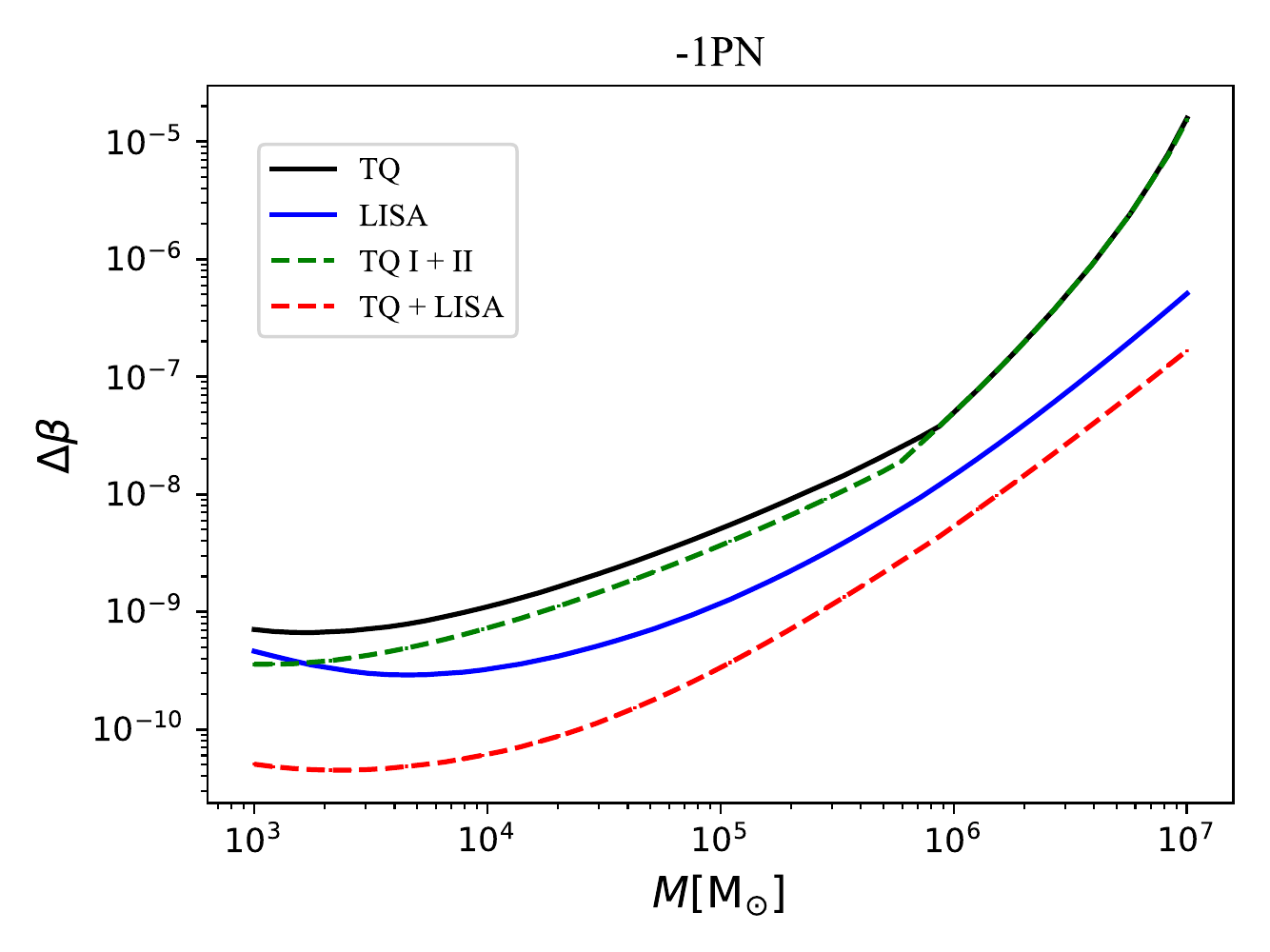}
\caption{Dependence of $\Delta\beta$ on $M$ at the $-1$ PN order for different detector configurations, with sources in the high mass range.}
\label{spmassdp-1}
\end{figure}

Different detectors are most sensitive to sources with different total masses, as such, the benefit of a detector network also varies with systems with different total masses. In this work, we study the benefit of a few detector networks, including the twin constellation configuration of TianQin (TQ I+II), the joint observation with TianQin and LISA (TQ + LISA), the multiband observation with TianQin and ET (TQ + ET), and the joint multiband observation with TianQin, LISA and ET (TQ + LISA + ET). The corresponding result can be found in \figurename{ \ref{stmassdp-1}} and \figurename{ \ref{spmassdp-1}}, and in \figurename{ \ref{fig.stmassall}} and \figurename{ \ref{fig.spmassall}}.

The two constellations in TQ I+II operate in a consecutive mode, with one starting to observe exactly when the other stops observation, at a rate of every three month. For sources in the low mass range, $M \in[10\mSun\,$, $10^3\mSun]\,$, TQ I+II in general can lead to 2$\sim$3 times improvement on the constraints on $\beta$. In the high mass range, $M \in[10^3\mSun, 10^7\mSun]\,$, TQ I+II improves over TianQin in a similar fashion, but the amount of improvement decreases with the source masses. There is no improvement of TQ I+II over TianQin for sources with a total mass greater than $9\times 10^5\mSun\,$. This is because such sources will merger in less than three months after it enters the TianQin frequency band at $10^{-4}$ Hz, and so TQ I+II is effectively in a one-constellation mode for such sources.

Due to the difference in their most sensitive frequency bands, TianQin and LISA have different constraining power at different PN orders and for sources with different masses: in the low mass range, $M \in[10\mSun\,$, $10^3\mSun]\,$, the constraints from TianQin is always better for all non-negative PN orders, and LISA starts to produce better constraints in more and more parameter range at the higher mass end for lower and lower PN orders; in the high mass range, $M \in[10^3\mSun, 10^7\mSun]\,$, the constraints from LISA is always better for all PN orders lower than 1PN, with TianQin being slightly better at the lower mass end for 2PN and 1.5PN. Except in the high mass range and for PN orders lower than $-2.5$PN, the the TQ + LISA network always outperforms the individual detectors by an appreciable amount, with the most significant improvement occurring at the 0PN order, by an amount of roughly three orders of magnitude for all source masses.

Both TianQin and ET can detect sources in the low mass range, $M \in[10\mSun\,$, $10^3\mSun]\,$. But there is significant difference in their capability in constraining $\beta\,$: while ET has a chance of being better at the lower mass end for PN orders no less than $-0.5$PN, TianQin becomes much better for all negative PN orders starting from $-1$PN. It is then interesting to note that the multiband observations of TQ + ET can always improve the constraints on $\beta$ by about $1\sim2$ orders of magnitude compared to individual detectors, even when ET is not able to place any competitive constraints by itself.

The effect of TQ + LISA + ET can be best seen comparing to those of TQ + ET and TQ + LISA. For the PN orders at 0PN and higher, TQ + ET is always better than TQ + LISA, and the constraints from TQ + LISA + ET mostly follow that of TQ + ET, becoming slightly better at the higher mass end. For the PN orders at $-0.5$PN and lower, TQ + LISA starts to become better than TQ + ET at the higher mass end, and the constraints from TQ + LISA + ET start to get aligned with that of TQ + LISA, being better than the latter by about 2$\sim$3 times.

\begin{table}[htbp]
\renewcommand{\arraystretch}{1.5}
\caption{A list of example sources.}
\label{tab:parameter}
\begin{tabular}{|m{1.6cm}<{\centering}|m{1.8cm}<{\centering}|m{1.6cm}<{\centering}|m{1.8cm}<{\centering}|}\hline
Type & $M$  & $q$  & $D_L$\\\hline
SBHB & $ 70  \mSun$  &  2  & 200 Mpc\\\hline
IMBHB  & $2\times10^3 \mSun$  & 2  & 2 Gpc\\\hline
MBHB & $2\times10^6 \mSun$  & 2  & 15 Gpc\\\hline
IMRI & $2\times10^5 \mSun$  & $2\times10^2$  & 1 Gpc\\\hline
EMRI & $2\times10^5 \mSun$  & $2\times10^4$  & 1 Gpc\\\hline
\end{tabular}
\end{table}

To get an idea on the specific numbers of the constraints on $\beta$, we use a set of example sources for TianQin \cite{Wang:2019ryf, Huang:2020rjf, Fan:2020zhy,Liu:2020eko, Liu:2021yoy} to calculate the constraints on $\beta$ for different detection scenarios. The sources are listed in \tablename{~\ref{tab:parameter}} and the constraints are listed in \tablename{~\ref{tab:result_SBHB}, \ref{tab:result_IMBHB}, \ref{tab:result_MBHB}, \ref{tab:result_IMRI}, \ref{tab:result_EMRI}}. Both \ac{EMRI} and \ac{IMRI} are included in the source list. Although the IMRphenomD and \ac{ppE} techniques are not suitable for such sources in principle, we use them to get an indicative idea of the level of expected constraints. For the detector configurations, apart from the ones that have already been considered, we also consider TianQin with three months of observation time (TQ\_3m) and TianQin with five years of observation time (TQ\_5y). The values in the tables are consistent with features already displayed in the relevant plots.

\section{Expected constraints on the EdGB and dCS theories}\label{sec:resultsII}

In this section, we discuss the projected constraints on the \ac{EdGB} and \ac{dCS} theories. Our main findings are the following.

\subsection{What kind of sources are the best for constraining the EdGB and dCS theories?}

In the \ac{EdGB} and \ac{dCS} theories, the leading order modifications to the inspiral waveform starts from the $-1$PN and $2$PN order, respectively, so some features of the source parameter dependence can already be read off from the corresponding plots in the last section. So in this section, we only present the detailed result on the dependence on the total mass $M$ and the symmetric mass ratio $\eta\,$, as these are the two most basic parameters.

The dependence of $\Delta\bar{\alpha}_{\rm EdGB}^2$ and $\Delta\bar{\alpha}_{\rm dCS}^2$  on $M$ and $\eta$ is plotted in \figurename{~\ref{fig.dcsedgb}}, together with the current bound on \ac{EdGB}, $\sqrt{\bar{\alpha}_{\rm EdGB}}\leq 1.3$ km, from the observations of GW200115 \cite{Lyu:2022gdr}, and the current best constraints on \ac{dCS}, $\sqrt{\bar{\alpha}_{\rm dCS}}\leq 8.7$ km, from the multi-messenger observations of GW170817 \cite{Silva:2020acr}.

For the dependence on the total mass, two important features can be noted. Firstly, the constraints on both theories improves monotonically as the total mass is lowered, making the low mass sources the better choice for constraining such theories. Secondly, the space-based detectors TianQin and LISA are better suited for constraining the \ac{EdGB} theory, while ET is better suited for constraining the \ac{dCS} theory.

For the dependence on the symmetric mass ratio, we have used three sources as examples:
\begin{itemize}
	\item \ac{SBHB}: $M=70{\mSun}\,$, $D_L= 200$$\rm Mpc\,$;
	\item \ac{IMBHB}: $M=2\times 10^3{\mSun}\,$, $D_L=2$ Gpc;
	\item \ac{MBHB}: $M=2\times10^6{\mSun}\,$, $D_L=15$ Gpc.
\end{itemize}
One can see that sources with smaller $\eta$ (corresponding to larger mass ratios) are better suited for constraining both the \ac{EdGB} and \ac{dCS} theories.

The current waveforms of the \ac{EdGB} and \ac{dCS} theories are derived in the small-coupling limit, thus the couplings in the theories have to satisfy the bound \cite{Perkins:2021mhb, Lyu:2022gdr},
\bea\bar{\alpha}_{\rm EdGB}^2,\bar{\alpha}_{\rm dCS}^2\lesssim\frac{m_2^4}{32 \pi}\,, \label{validity}\eea
where $m_2$ is the mass of the minor. This bound has also been plotted in the upper and middle panels in \figurename~\ref{fig.dcsedgb}, and only results in the regions below the bound is considered reliable. One can see that, although the \ac{EdGB} theory can get reliable constraints, all results for the \ac{dCS} theory are above the bound and so cannot be taken too seriously.

We have also used the example sources in \tablename{~\ref{tab:parameter}} to calculate the constraints on $\bar\alpha_{\rm EdGB}$ and $\bar\alpha_{\rm dCS}$ for different detector configurations. We find that, with TianQin, there is chance to use \ac{SBHB} to constrain $\sqrt{\bar{\alpha}_{\rm EdGB}}$ to the level $\mathcal{O}(10^{\rm -1}$) km, which is about one order of magnitude improvement over the current bound.

\subsection{Will the amplitude correction affect the constraints on the EdGB and dCS theories?}\label{sec:amp}

A laser interferometric \ac{GW} detector can measure the \ac{GW} phase much better than its amplitude, and so the majority work of testing \ac{GR} does not involve the amplitude correction, which is characterized by the $\alpha$ parameter in (\ref{eq:waveform}).

\begin{figure*}
\centering
\includegraphics[width=1\textwidth]{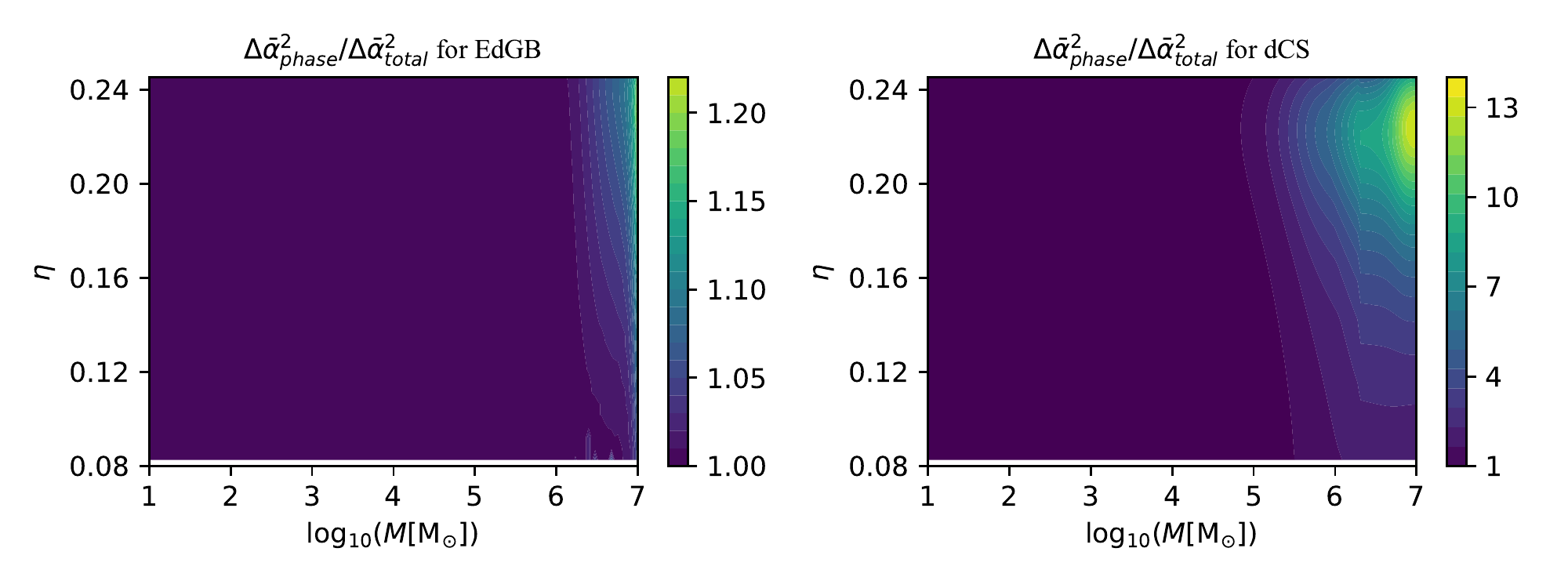}
\caption{Effect of amplitude correction on the constraints on \ac{EdGB} (left) and \ac{dCS} (right) theories.}
\label{fig.aip}
\end{figure*}

The effect of the amplitude correction is illustrated in \figurename{~\ref{fig.aip}, where $\Delta\bar{\alpha}^2_{phase}$ represents the constraints on the coupling constants $\bar\alpha_{\rm EdGB}$ or $\bar\alpha_{\rm dCS}\,$, in the case when only the phase correction is considered (i.e., assuming $\alpha=0$), while $\Delta\bar{\alpha}^2_{total}$ represents the corresponding constraint when both the phase correction and the amplitude correction are considered. One can see that for sources with a total mass $M>10^6\mSun$ the amplitude correction can bring some improvement on the constraints. However, since such massive sources cannot yield competitive constraints on these two theories compared to sources in the low mass range, one can conclude that the amplitude correction is not important for constraining the \ac{EdGB} or \ac{dCS} theory as a whole.

\section{Summary}\label{sec:summary}

In this paper, we have carried out a detailed study of the prospect of using TianQin to do a broad test of \ac{GR}, by using the inspiral signal of black hole binaries and the \ac{ppE} waveform. We have compared the capability of TianQin to two other important detectors, \ac{ET} and LISA, and have studied the scientific potential of detector networks made of TianQin and these detectors. We have also studied the constraints on specific theories such as the \ac{EdGB} and \ac{dCS} theories.

We mainly focus on the constraints on the phase correction parameter $\beta$ in the \ac{ppE} waveform (\ref{eq:waveform}), at PN orders ranging from $-4$PN to 2PN. For the effect of different parameters on the constraints on $\beta$, we have mainly focused on the total mass $M\,$, the symmetric mass ratio $\eta\,$, the total observation time $T_{ob}$ and the missed observation time $T_m$ (this last one is for TianQin only). We find that all these parameters have notable effect on the expected constraints. Depending on the chosen PN order, there can be orders of magnitude change in the constraints. For example, at the $-4$PN order, the difference can be more than eight orders of magnitude when the total mass is varied in the high mass range, $M \in[10^3\mSun\,$, $10^7\mSun]\,$. 

The missed observation time of TianQin $T_m$ can also make a big impact. In the worst case scenarios when all three months of data is lost right before \ac{ISCO}, the \ac{MBHB} signals will become hardly detectable, while there can still be partial \ac{SNR} left for the  \ac{SBHB} and \ac{IMBHB} signals. For two example sources considered in this paper, 40\% and 76\% of \acp{SNR} are found to be left for \ac{SBHB} and \ac{IMBHB}, respectively. The constraints on $\beta$ can be worsened by about $1\sim2$ orders of magnitude due to the presence of $T_m\,$.

We have compared the capability of TianQin to other detectors. For example, compared with the results of GW150914 \cite{Yunes:2016jcc}, TianQin can improve the constraints by several orders of magnitude for different PN corrections, e.g., nearly $13$ orders of magnitude at $-4$PN (with \acp{SBHB}) and nearly 3 orders at 2PN (with \acp{IMRI}). Compared to ET, TianQin is better below the $-1$PN order, and for corrections above the $-0.5$PN order, TianQin is still competitive near the large mass end of the low mass range $[10\mSun\,,\,10^3\mSun]\,$. Compared to LISA, TianQin can be competitive in the lower mass end as the PN order is increased. For example, at the $-4$PN order, LISA is always superior for sources more massive than about $30\mSun\,$, while at the $2$PN order, TianQin becomes competitive for sources less massive than about $10^4\mSun\,$.

We have considered multiple detector configurations involving TianQin, LISA and ET. We find that:
\begin{enumerate}
\item TQ I+II can improve the constraints on $\beta$ by about $2\sim3$ times, comparing to TianQin alone, in the low mass range, while in the high mass range the improvement diminishes as the total mass increases;
\item Except in the high mass range and for PN orders lower than $-2.5$PN, the TQ + LISA network always outperforms the individual detectors by an appreciable amount, with the most significant improvement occurring at the 0PN order, by an amount of roughly three orders of magnitude for all source masses;
\item The multiband observations of TQ + ET can always improve the constraints on $\beta$ by about $1\sim2$ orders of magnitude compared to individual detectors, even when ET is not able to place any competitive constraints by itself;
\item TQ + LISA + ET is always better than TQ + ET and TQ + LISA, and the improvement can reach 2$\sim$3 times.
\end{enumerate}

We have also considered the constraints on specific theories such as the \ac{EdGB} and \ac{dCS} theories. We find that reliable constraints can be placed on the \ac{EdGB} theory. If TianQin can detect a low mass source with total mass at the order $\mathcal{O}(10)\mSun$ at about $D_L\approx 200 {\rm~Mpc}\,$, then one can get a constraint on the \ac{EdGB} theory at the order $\sqrt{\bar{\alpha}_{\rm EdGB}}<\mathcal{O}(10^{-1}$ km), which is about an order of magnitude improvement over the current best bound. For the \ac{dCS} theory, no reliable constraints can be obtained with the detectors considered in this paper, due to a lack of reliable waveform.

\begin{acknowledgments}
The authors thank Yi-Ming Hu for useful discussions and Kent Yagi for the helpful communication. This work has been supported  by the Guangdong Basic and Applied Basic Research Foundation(Grant No. 2021A1515010319), the Guangdong Major Project of Basic and Applied Basic Research (Grant No. 2019B030302001).
\end{acknowledgments}

\bibliographystyle{apsrev4-1}
\bibliography{reference}

\begin{widetext}
\begin{center}

\begin{table}[htbp]
	\renewcommand{\arraystretch}{1.5}
	\footnotesize
	\caption{Constraints on $\beta_{\rm ppE}$ for different detector configurations with \ac{SBHB}.}
	\label{tab:result_SBHB}
	\begin{tabular}{|m{1.5cm}<{\centering}|m{1cm}<{\centering}|m{1cm}<{\centering}|m{1cm}<{\centering}|m{1cm}<{\centering}|m{1cm}<{\centering}|m{1cm}<{\centering}|m{1cm}<{\centering}|m{1cm}<{\centering}|m{1cm}<{\centering}|m{1cm}<{\centering}|m{1cm}<{\centering}|m{1cm}<{\centering}|m{1cm}<{\centering}|}
		\hline
		 PN order & $-4$ & $-3.5$ & $-3$ & $-2.5$ & $-2$ & $-1.5$ & $-1$ & $-0.5$ & $0$ & $0.5$ & $1$ & $1.5$ & $2$ \\
		\hline
		TQ & $2.4\times10^{-21}$ & $1.1\times10^{-19}$ & $4.7\times10^{-18}$ & $2.1\times10^{-16}$ & $1.0\times10^{-14}$ & $5.2\times10^{-13}$ & $3.0\times10^{-11}$ & $2.4\times10^{-9}$ & $1.6\times10^{-5}$ & $4.1\times10^{-6}$ & $9.0\times10^{-5}$ & $3.0\times10^{-3}$ & $1.5\times10^{-1}$\\\hline
		TQ\_3m & $9.5\times10^{-21}$ & $3.8\times10^{-19}$ & $1.5\times10^{-17}$ & $6.5\times10^{-16}$ & $2.9\times10^{-14}$ & $1.3\times10^{-12}$ & $7.3\times10^{-11}$ & $5.3\times10^{-9}$ & $2.8\times10^{-5}$ & $7.7\times10^{-6}$ & $1.6\times10^{-4}$ & $4.9\times10^{-3}$ & $2.3\times10^{-1}$\\\hline
		TQ\_5y & $1.1\times10^{-22}$ & $5.7\times10^{-21}$ & $3.1\times10^{-19}$ & $1.7\times10^{-17}$ & $9.9\times10^{-16}$ & $6.1\times10^{-14}$ & $4.3\times10^{-12}$ & $4.1\times10^{-10}$ & $4.3\times10^{-6}$ & $9.8\times10^{-7}$ & $2.6\times10^{-5}$ & $1.0\times10^{-3}$ & $6.0\times10^{-2}$\\\hline
		TQ I+II & $6.4\times10^{-22}$ & $3.0\times10^{-20}$ & $1.5\times10^{-18}$ & $7.2\times10^{-17}$ & $3.7\times10^{-15}$ & $2.1\times10^{-13}$ & $1.3\times10^{-11}$ & $1.1\times10^{-9}$ & $9.1\times10^{-6}$ & $2.2\times10^{-6}$ & $5.2\times10^{-5}$ & $1.9\times10^{-3}$ & $1.0\times10^{-1}$\\\hline
		LISA & $1.7\times10^{-21}$ & $8.3\times10^{-20}$ & $4.1\times10^{-18}$ & $2.1\times10^{-16}$ & $1.1\times10^{-14}$ & $6.6\times10^{-13}$ & $4.3\times10^{-11}$ & $3.8\times10^{-9}$ & $3.6\times10^{-5}$ & $8.5\times10^{-6}$ & $2.1\times10^{-4}$ & $8.2\times10^{-3}$ & $4.8\times10^{-1}$\\\hline
		ET & $1.4\times10^{-15}$ & $1.5\times10^{-14}$ & $1.6\times10^{-13}$ & $1.8\times10^{-12}$ & $2.0\times10^{-11}$ & $2.4\times10^{-10}$ & $3.3\times10^{-9}$ & $6.0\times10^{-8}$ & $5.7\times10^{-6}$ & $4.2\times10^{-6}$ & $2.1\times10^{-5}$ & $1.5\times10^{-4}$ & $1.7\times10^{-3}$\\\hline
		TQ + LISA & $3.7\times10^{-22}$ & $1.6\times10^{-20}$ & $6.9\times10^{-19}$ & $3.0\times10^{-17}$ & $1.3\times10^{-15}$ & $5.7\times10^{-14}$ & $2.5\times10^{-12}$ & $1.1\times10^{-10}$ & $5.1\times10^{-9}$ & $2.4\times10^{-7}$ & $1.2\times10^{-5}$ & $6.5\times10^{-4}$ & $4.8\times10^{-2}$\\\hline
		TQ + ET & $5.4\times10^{-22}$ & $2.1\times10^{-20}$ & $7.9\times10^{-19}$ & $3.0\times10^{-17}$ & $1.2\times10^{-15}$ & $4.5\times10^{-14}$ & $1.7\times10^{-12}$ & $6.5\times10^{-11}$ & $2.5\times10^{-9}$ & $9.4\times10^{-8}$ & $2.9\times10^{-6}$ & $4.5\times10^{-5}$ & $6.6\times10^{-4}$\\\hline
		TQ + LISA + ET & $2.4\times10^{-22}$ & $1.0\times10^{-20}$ & $4.2\times10^{-19}$ & $1.7\times10^{-17}$ & $7.1\times10^{-16}$ & $2.9\times10^{-14}$ & $1.2\times10^{-12}$ & $4.8\times10^{-11}$ & $1.9\times10^{-9}$ & $7.7\times10^{-8}$ & $2.6\times10^{-6}$ & $4.5\times10^{-5}$ & $6.6\times10^{-4}$\\\hline
	\end{tabular}
\end{table}

\begin{table}[htbp]
\renewcommand{\arraystretch}{1.5}
\footnotesize
\caption{Constraints on $\beta_{\rm ppE}$ for different detector configurations with IMBHB.}
\label{tab:result_IMBHB}
\begin{tabular}{|m{1.5cm}<{\centering}|m{1cm}<{\centering}|m{1cm}<{\centering}|m{1cm}<{\centering}|m{1cm}<{\centering}|m{1cm}<{\centering}|m{1cm}<{\centering}|m{1cm}<{\centering}|m{1cm}<{\centering}|m{1cm}<{\centering}|m{1cm}<{\centering}|m{1cm}<{\centering}|m{1cm}<{\centering}|m{1cm}<{\centering}|}
\hline
 PN order & $-4$ & $-3.5$ & $-3$ & $-2.5$ & $-2$ & $-1.5$ & $-1$ & $-0.5$ & $0$ & $0.5$ & $1$ & $1.5$ & $2$\\ \hline
TQ & $1.1\times10^{-19}$ & $3.2\times10^{-18}$ & $8.9\times10^{-17}$ & $2.6\times10^{-15}$ & $7.7\times10^{-14}$ & $2.5\times10^{-12}$ & $8.9\times10^{-11}$ & $4.2\times10^{-9}$ & $7.1\times10^{-6}$ & $2.7\times10^{-6}$ & $3.4\times10^{-5}$ & $6.7\times10^{-4}$ & $1.9\times10^{-2}$\\\hline
TQ\_3m & $3.4\times10^{-19}$ & $8.7\times10^{-18}$ & $2.2\times10^{-16}$ & $6.0\times10^{-15}$ & $1.7\times10^{-13}$ & $5.0\times10^{-12}$ & $1.7\times10^{-10}$ & $7.5\times10^{-9}$ & $1.1\times10^{-5}$ & $4.1\times10^{-6}$ & $5.0\times10^{-5}$ & $9.0\times10^{-4}$ & $2.4\times10^{-2}$\\\hline
TQ\_5y & $1.2\times10^{-20}$ & $4.0\times10^{-19}$ & $1.3\times10^{-17}$ & $4.6\times10^{-16}$ & $1.6\times10^{-14}$ & $6.1\times10^{-13}$ & $2.6\times10^{-11}$ & $1.5\times10^{-9}$ & $3.5\times10^{-6}$ & $1.2\times10^{-6}$ & $1.8\times10^{-5}$ & $3.9\times10^{-4}$ & $1.2\times10^{-2}$\\\hline
TQ I+II & $4.2\times10^{-20}$ & $1.3\times10^{-18}$ & $3.8\times10^{-17}$ & $1.2\times10^{-15}$ & $3.8\times10^{-14}$ & $1.3\times10^{-12}$ & $5.1\times10^{-11}$ & $2.6\times10^{-9}$ & $5.1\times10^{-6}$ & $1.9\times10^{-6}$ & $2.5\times10^{-5}$ & $5.1\times10^{-4}$ & $1.5\times10^{-2}$\\\hline
LISA & $2.7\times10^{-20}$ & $8.5\times10^{-19}$ & $2.7\times10^{-17}$ & $8.8\times10^{-16}$ & $3.0\times10^{-14}$ & $1.1\times10^{-12}$ & $4.5\times10^{-11}$ & $2.5\times10^{-9}$ & $6.2\times10^{-6}$ & $2.1\times10^{-6}$ & $3.3\times10^{-5}$ & $7.5\times10^{-4}$ & $2.6\times10^{-2}$\\\hline
TQ + LISA & $1.2\times10^{-20}$ & $3.4\times10^{-19}$ & $9.4\times10^{-18}$ & $2.6\times10^{-16}$ & $7.5\times10^{-15}$ & $2.1\times10^{-13}$ & $6.1\times10^{-12}$ & $1.8\times10^{-10}$ & $5.2\times10^{-9}$ & $1.6\times10^{-7}$ & $5.1\times10^{-6}$ & $1.9\times10^{-4}$ & $9.1\times10^{-3}$\\\hline
\end{tabular}
\end{table}

\begin{table}[htbp]
\renewcommand{\arraystretch}{1.5}
\footnotesize
\caption{Constraints on $\beta_{\rm ppE}$ for different detector configurations with \ac{MBHB}.}
\label{tab:result_MBHB}
\begin{tabular}{|m{1.5cm}<{\centering}|m{1cm}<{\centering}|m{1cm}<{\centering}|m{1cm}<{\centering}|m{1cm}<{\centering}|m{1cm}<{\centering}|m{1cm}<{\centering}|m{1cm}<{\centering}|m{1cm}<{\centering}|m{1cm}<{\centering}|m{1cm}<{\centering}|m{1cm}<{\centering}|m{1cm}<{\centering}|m{1cm}<{\centering}|}
\hline
 PN order & $-4$ & $-3.5$ & $-3$ & $-2.5$ & $-2$ & $-1.5$ & $-1$ & $-0.5$ & $0$ & $0.5$ & $1$ & $1.5$ & $2$\\\hline
TQ & $2.0\times10^{-13}$ & $1.8\times10^{-12}$ & $1.7\times10^{-11}$ & $1.7\times10^{-10}$ & $1.7\times10^{-9}$ & $1.7\times10^{-8}$ & $2.0\times10^{-7}$ & $3.0\times10^{-6}$ & $8.8\times10^{-4}$ & $1.9\times10^{-4}$ & $8.1\times10^{-4}$ & $5.0\times10^{-3}$ & $4.7\times10^{-2}$\\\hline
TQ\_3m & $2.0\times10^{-13}$ & $1.8\times10^{-12}$ & $1.7\times10^{-11}$ & $1.7\times10^{-10}$ & $1.7\times10^{-9}$ & $1.7\times10^{-8}$ & $2.0\times10^{-7}$ & $3.0\times10^{-6}$ & $8.8\times10^{-4}$ & $1.9\times10^{-4}$ & $8.1\times10^{-4}$ & $5.0\times10^{-3}$ & $4.7\times10^{-2}$\\\hline
TQ\_5y & $2.0\times10^{-13}$ & $1.8\times10^{-12}$ & $1.7\times10^{-11}$ & $1.7\times10^{-10}$ & $1.7\times10^{-9}$ & $1.7\times10^{-8}$ & $2.0\times10^{-7}$ & $3.0\times10^{-6}$ & $8.8\times10^{-4}$ & $1.9\times10^{-4}$ & $8.1\times10^{-4}$ & $5.0\times10^{-3}$ & $4.7\times10^{-2}$\\\hline
TQ I+II & $2.0\times10^{-13}$ & $1.8\times10^{-12}$ & $1.7\times10^{-11}$ & $1.7\times10^{-10}$ & $1.7\times10^{-9}$ & $1.7\times10^{-8}$ & $2.0\times10^{-7}$ & $3.0\times10^{-6}$ & $8.8\times10^{-4}$ & $1.9\times10^{-4}$ & $8.1\times10^{-4}$ & $5.0\times10^{-3}$ & $4.7\times10^{-2}$\\\hline
LISA & $1.5\times10^{-14}$ & $1.7\times10^{-13}$ & $1.9\times10^{-12}$ & $2.2\times10^{-11}$ & $2.5\times10^{-10}$ & $3.0\times10^{-9}$ & $3.9\times10^{-8}$ & $6.6\times10^{-7}$ & $1.9\times10^{-4}$ & $4.7\times10^{-5}$ & $2.0\times10^{-4}$ & $1.3\times10^{-3}$ & $1.2\times10^{-2}$\\\hline
TQ + LISA & $1.3\times10^{-14}$ & $1.4\times10^{-13}$ & $1.5\times10^{-12}$ & $1.6\times10^{-11}$ & $1.6\times10^{-10}$ & $1.6\times10^{-9}$ & $1.4\times10^{-8}$ & $1.2\times10^{-7}$ & $1.0\times10^{-6}$ & $7.8\times10^{-6}$ & $6.2\times10^{-5}$ & $5.4\times10^{-4}$ & $6.2\times10^{-3}$\\\hline
\end{tabular}
\end{table}

\begin{table}[htbp]
\renewcommand{\arraystretch}{1.5}
\footnotesize
\caption{Constraints on $\beta_{\rm ppE}$ for different detector configurations with \ac{IMRI}.}
\label{tab:result_IMRI}
\begin{tabular}{|m{1.5cm}<{\centering}|m{1cm}<{\centering}|m{1cm}<{\centering}|m{1cm}<{\centering}|m{1cm}<{\centering}|m{1cm}<{\centering}|m{1cm}<{\centering}|m{1cm}<{\centering}|m{1cm}<{\centering}|m{1cm}<{\centering}|m{1cm}<{\centering}|m{1cm}<{\centering}|m{1cm}<{\centering}|m{1cm}<{\centering}|}
\hline
 PN order & $-4$ & $-3.5$ & $-3$ & $-2.5$ & $-2$ & $-1.5$ & $-1$ & $-0.5$ & $0$ & $0.5$ & $1$ & $1.5$ & $2$\\\hline
TQ & $1.5\times10^{-18}$ & $3.0\times10^{-17}$ & $6.0\times10^{-16}$ & $1.3\times10^{-14}$ & $2.7\times10^{-13}$ & $6.3\times10^{-12}$ & $1.6\times10^{-10}$ & $5.6\times10^{-9}$ & $4.1\times10^{-6}$ & $1.9\times10^{-6}$ & $1.8\times10^{-5}$ & $2.6\times10^{-4}$ & $5.6\times10^{-3}$\\\hline
TQ\_3m & $3.5\times10^{-18}$ & $6.6\times10^{-17}$ & $1.3\times10^{-15}$ & $2.4\times10^{-14}$ & $4.9\times10^{-13}$ & $1.1\times10^{-11}$ & $2.6\times10^{-10}$ & $8.5\times10^{-9}$ & $6.2\times10^{-6}$ & $2.6\times10^{-6}$ & $2.4\times10^{-5}$ & $3.3\times10^{-4}$ & $6.9\times10^{-3}$\\\hline
TQ\_5y & $2.1\times10^{-19}$ & $5.0\times10^{-18}$ & $1.2\times10^{-16}$ & $3.0\times10^{-15}$ & $7.6\times10^{-14}$ & $2.1\times10^{-12}$ & $6.2\times10^{-11}$ & $2.4\times10^{-9}$ & $2.0\times10^{-6}$ & $1.0\times10^{-6}$ & $1.1\times10^{-5}$ & $1.7\times10^{-4}$ & $4.0\times10^{-3}$\\\hline
TQ I+II & $6.3\times10^{-19}$ & $1.4\times10^{-17}$ & $3.0\times10^{-16}$ & $6.7\times10^{-15}$ & $1.6\times10^{-13}$ & $3.8\times10^{-12}$ & $1.1\times10^{-10}$ & $3.8\times10^{-9}$ & $2.9\times10^{-6}$ & $1.4\times10^{-6}$ & $1.4\times10^{-5}$ & $2.1\times10^{-4}$ & $4.8\times10^{-3}$\\\hline
LISA & $1.7\times10^{-19}$ & $3.8\times10^{-18}$ & $8.6\times10^{-17}$ & $2.0\times10^{-15}$ & $4.8\times10^{-14}$ & $1.2\times10^{-12}$ & $3.5\times10^{-11}$ & $1.3\times10^{-9}$ & $1.0\times10^{-6}$ & $5.6\times10^{-7}$ & $5.9\times10^{-6}$ & $9.6\times10^{-5}$ & $2.3\times10^{-3}$\\\hline
TQ + LISA & $1.0\times10^{-19}$ & $2.1\times10^{-18}$ & $4.2\times10^{-17}$ & $8.5\times10^{-16}$ & $1.7\times10^{-14}$ & $3.4\times10^{-13}$ & $6.9\times10^{-12}$ & $1.4\times10^{-10}$ & $2.8\times10^{-9}$ & $5.7\times10^{-8}$ & $1.2\times10^{-6}$ & $2.8\times10^{-5}$ & $8.7\times10^{-4}$\\\hline
\end{tabular}
\end{table}

\begin{table}[htbp]
\renewcommand{\arraystretch}{1.5}
\footnotesize
\caption{Constraints on $\beta_{\rm ppE}$ for different detector configurations with \ac{EMRI}.}
\label{tab:result_EMRI}
\begin{tabular}{|m{1.5cm}<{\centering}|m{1cm}<{\centering}|m{1cm}<{\centering}|m{1cm}<{\centering}|m{1cm}<{\centering}|m{1cm}<{\centering}|m{1cm}<{\centering}|m{1cm}<{\centering}|m{1cm}<{\centering}|m{1cm}<{\centering}|m{1cm}<{\centering}|m{1cm}<{\centering}|m{1cm}<{\centering}|m{1cm}<{\centering}|}
\hline
 PN order & $-4$ & $-3.5$ & $-3$ & $-2.5$ & $-2$ & $-1.5$ & $-1$ & $-0.5$ & $0$ & $0.5$ & $1$ & $1.5$ & $2$\\\hline
TQ & $8.4\times10^{-20}$ & $2.7\times10^{-18}$ & $8.8\times10^{-17}$ & $3.0\times10^{-15}$ & $1.1\times10^{-13}$ & $4.1\times10^{-12}$ & $1.8\times10^{-10}$ & $1.1\times10^{-8}$ & $4.2\times10^{-6}$ & $1.2\times10^{-5}$ & $2.1\times10^{-4}$ & $5.8\times10^{-3}$ & $2.4\times10^{-1}$\\\hline
TQ\_3m & $3.4\times10^{-19}$ & $1.0\times10^{-17}$ & $3.1\times10^{-16}$ & $9.9\times10^{-15}$ & $3.3\times10^{-13}$ & $1.2\times10^{-11}$ & $4.9\times10^{-10}$ & $2.7\times10^{-8}$ & $4.6\times10^{-6}$ & $3.2\times10^{-5}$ & $5.1\times10^{-4}$ & $1.3\times10^{-2}$ & $5.3\times10^{-1}$\\\hline
TQ\_5y & $5.5\times10^{-21}$ & $2.1\times10^{-19}$ & $7.8\times10^{-18}$ & $3.1\times10^{-16}$ & $1.3\times10^{-14}$ & $5.6\times10^{-13}$ & $2.8\times10^{-11}$ & $1.9\times10^{-9}$ & $2.6\times10^{-6}$ & $2.6\times10^{-6}$ & $5.1\times10^{-5}$ & $1.6\times10^{-3}$ & $7.3\times10^{-2}$\\\hline
TQ I+II & $2.4\times10^{-20}$ & $8.3\times10^{-19}$ & $2.9\times10^{-17}$ & $1.0\times10^{-15}$ & $3.9\times10^{-14}$ & $1.6\times10^{-12}$ & $7.5\times10^{-11}$ & $4.7\times10^{-9}$ & $3.7\times10^{-6}$ & $5.8\times10^{-6}$ & $1.1\times10^{-4}$ & $3.0\times10^{-4}$ & $1.3\times10^{-1}$\\\hline
LISA & $1.7\times10^{-20}$ & $5.8\times10^{-19}$ & $2.1\times10^{-17}$ & $7.6\times10^{-16}$ & $2.9\times10^{-14}$ & $1.2\times10^{-12}$ & $5.8\times10^{-11}$ & $3.7\times10^{-9}$ & $4.3\times10^{-6}$ & $4.7\times10^{-6}$ & $8.8\times10^{-5}$ & $2.6\times10^{-3}$ & $1.2\times10^{-1}$\\\hline
TQ + LISA & $5.8\times10^{-21}$ & $1.8\times10^{-19}$ & $5.3\times10^{-18}$ & $1.6\times10^{-16}$ & $5.0\times10^{-15}$ & $1.6\times10^{-13}$ & $4.9\times10^{-12}$ & $1.6\times10^{-10}$ & $5.3\times10^{-9}$ & $1.9\times10^{-7}$ & $7.0\times10^{-6}$ & $3.1\times10^{-4}$ & $1.8\times10^{-2}$\\\hline
\end{tabular}
\end{table}

\begin{table}[htbp]
\footnotesize
\renewcommand{\arraystretch}{1.5}
\caption{Constraints on \ac{EdGB} with example sources and for different detector configurations. ``Validity" means the bound imposed by (\ref{validity}), and a result is considered reliable only when it is below the indicated bound.}
\label{tab:result_EdGB}
\begin{tabular}{|m{1.5cm}<{\centering}|m{1.5cm}<{\centering}|m{1.3cm}<{\centering}|m{1.3cm}<{\centering}|m{1.3cm}<{\centering}|m{1.3cm}<{\centering}|m{1.3cm}<{\centering}|m{1.3cm}<{\centering}|m{1.3cm}<{\centering}|m{1.3cm}<{\centering}|m{1.3cm}<{\centering}|}\hline
Parameter & Source (Validity) & TQ &  TQ\_3m  &  TQ\_5y &  TQ I+II &  LISA  &  ET  &  TQ + LISA  &  TQ + ET  & TQ + LISA + ET  \\\hline
\multirow{7}{*}{ $\displaystyle\frac{\Delta\alpha_{\rm EdGB}^2}{[\rm km^4]}$} & \ac{SBHB} ($1.4\times10^4$) & $4.0\times10^{-3}$  &  $9.5\times10^{-3}$  & $5.8\times10^{-4}$  & $1.7\times10^{-3}$ & $4.7\times10^{-3}$ & $5.4\times10^{-1}$ & $3.2\times10^{-4}$ & $2.2\times10^{-4}$ & $1.4\times10^{-4}$\\\cline{2-11}
& \ac{IMBHB} ($9.4\times10^9$)  & $7.7\times10^{3}$  &  $1.5\times10^{4}$  & $2.2\times10^{3}$ & $4.4\times10^{3}$ & $3.4\times10^{3}$ & \_ & $5.3\times10^{2}$ & \_ & \_\\\cline{2-11}
& \ac{MBHB} ($9.4\times10^{21}$) & $1.7\times10^{19} $  & $1.7\times10^{19} $   & $1.7\times10^{19} $  & $1.7\times10^{19} $  & $3.0\times10^{18} $ & \_ & $1.2\times10^{18} $  & \_ & \_ \\\cline{2-11}
& \ac{IMRI} ($9.4\times10^9$) & $1.8\times10^{5}$  & $3.0\times10^{5}$  & $7.0\times 10^{4}$ & $1.8\times10^{5}$ & $3.3\times10^{4}$ & \_ & $7.4\times10^{3}$ & \_ & \_\\\cline{2-11}
& \ac{EMRI} ($9.4\times10^5$) & $1.3\times10^{-2}$ & $3.6\times10^{-2}$ & $2.0\times10^{-3}$ & $5.5\times10^{-3}$ & $2.2\times10^{-3}$ & \_ & $2.7\times10^{-4}$ & \_ & \_\\\hline
\end{tabular}
\end{table}

\begin{table}[htbp]
\footnotesize
\renewcommand{\arraystretch}{1.5}
\caption{Constraints on \ac{dCS} with example sources and for different detector configurations. ``Validity" means the bound imposed by (\ref{validity}), and a result is considered reliable only when it is below the indicated bound.}
\label{tab:result_dCS}
\begin{tabular}{|m{1.5cm}<{\centering}|m{1.5cm}<{\centering}|m{1.3cm}<{\centering}|m{1.3cm}<{\centering}|m{1.3cm}<{\centering}|m{1.3cm}<{\centering}|m{1.3cm}<{\centering}|m{1.3cm}<{\centering}|m{1.3cm}<{\centering}|m{1.3cm}<{\centering}|m{1.3cm}<{\centering}|}\hline
Parameter & Source (Validity) & TQ &  TQ\_3m  &  TQ\_5y &  TQ I+II &  LISA &  ET & TQ + LISA & TQ + ET & TQ + LISA + ET  \\\hline
\multirow{7}{*}{ $\displaystyle\frac{\Delta\alpha_{\rm dCS}^2}{[\rm km^4]}$ } & \ac{SBHB} ($1.4\times10^4$) & $8.6\times10^6$  &  $1.3\times10^7$ & $3.3\times10^6$ & $5.7\times10^6$ & $2.7\times10^{7}$ & $9.2\times10^4$ & $2.5\times10^6$ & $3.7\times10^4$ & $3.7\times10^4$\\\cline{2-11}
& \ac{IMBHB} ($9.4\times10^9$) & $7.8\times10^{12}$  &  $1.1\times10^{13}$  & $4.5\times10^{12}$ & $6.0\times10^{12}$ & $1.2\times10^{13}$ & \_ & $3.4\times10^{11}$ & \_ & \_\\\cline{2-11}
& \ac{MBHB} ($9.4\times10^{21}$) & $1.6\times10^{24} $  & $1.6\times10^{24} $   & $1.6\times10^{24} $  & $1.6\times10^{24} $  & $4.2\times10^{23} $ & \_ & $2.2\times10^{23} $  & \_ & \_ \\\cline{2-11}
& \ac{IMRI} ($9.4\times10^9$) & $5.9\times 10^{13}$  & $7.2\times10^{13}$  & $4.1\times10^{13}$ & $5.0\times10^{13}$ & $2.4\times10^{13}$ & \_ & $9.0\times10^{12}$ & \_ & \_\\\cline{2-11}
& \ac{EMRI} ($9.4\times10^5$) & $6.4\times10^{9}$ & $1.4\times10^{10}$ & $1.9\times10^{9}$ & $3.5\times^{9}$ & $3.0\times10^{9}$ & \_ & $5.0\times10^{8}$ & \_ & \_\\\hline
\end{tabular}
\end{table}

\begin{figure*}
	\centering
	\includegraphics[width=1\textwidth]{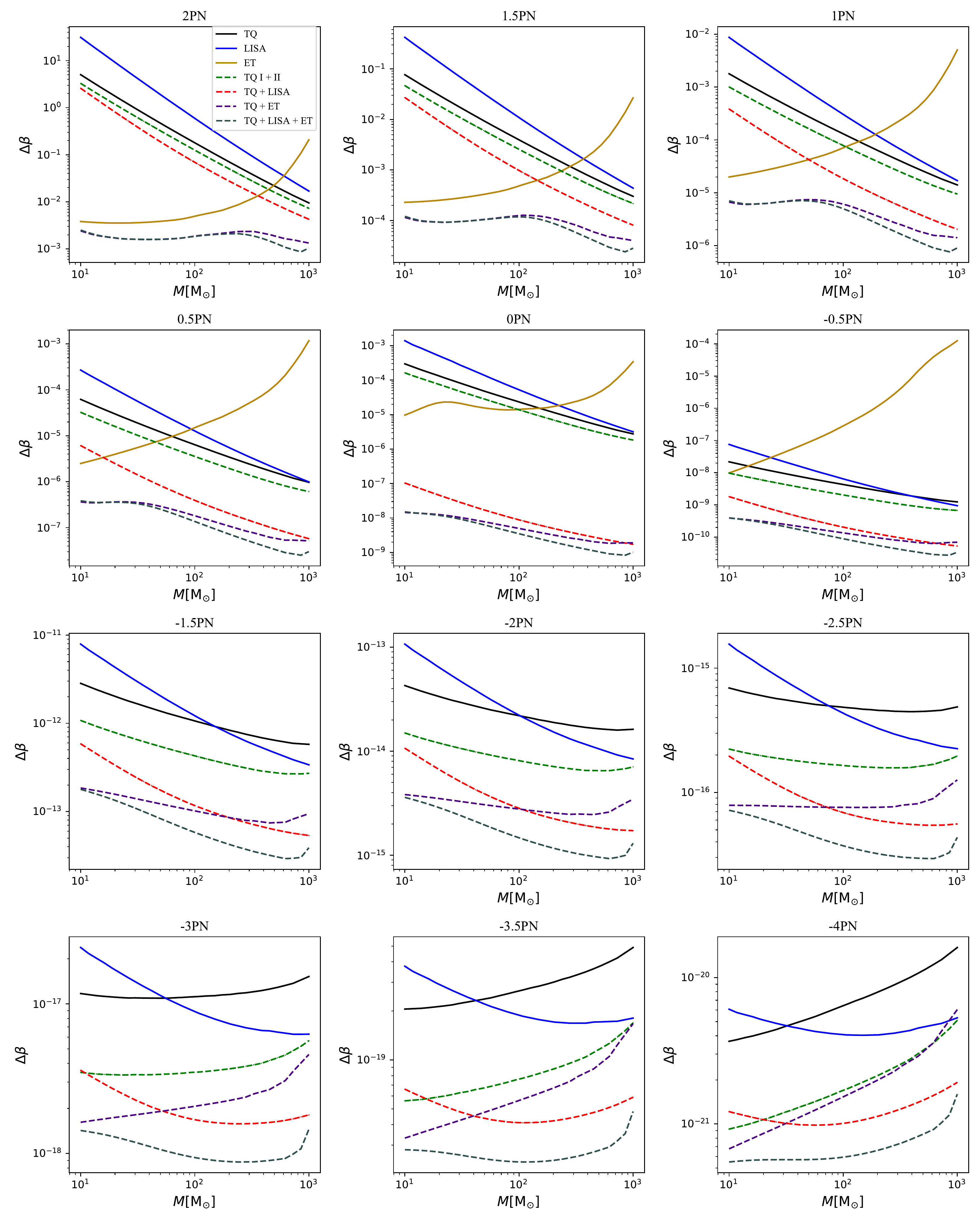}
	\caption{Dependence of $\Delta\beta$ on the total mass in the low mass range at different PN orders for different detector configurations.}
	\label{fig.stmassall}
\end{figure*}

\begin{figure*}
	\centering
	\includegraphics[width=1\textwidth]{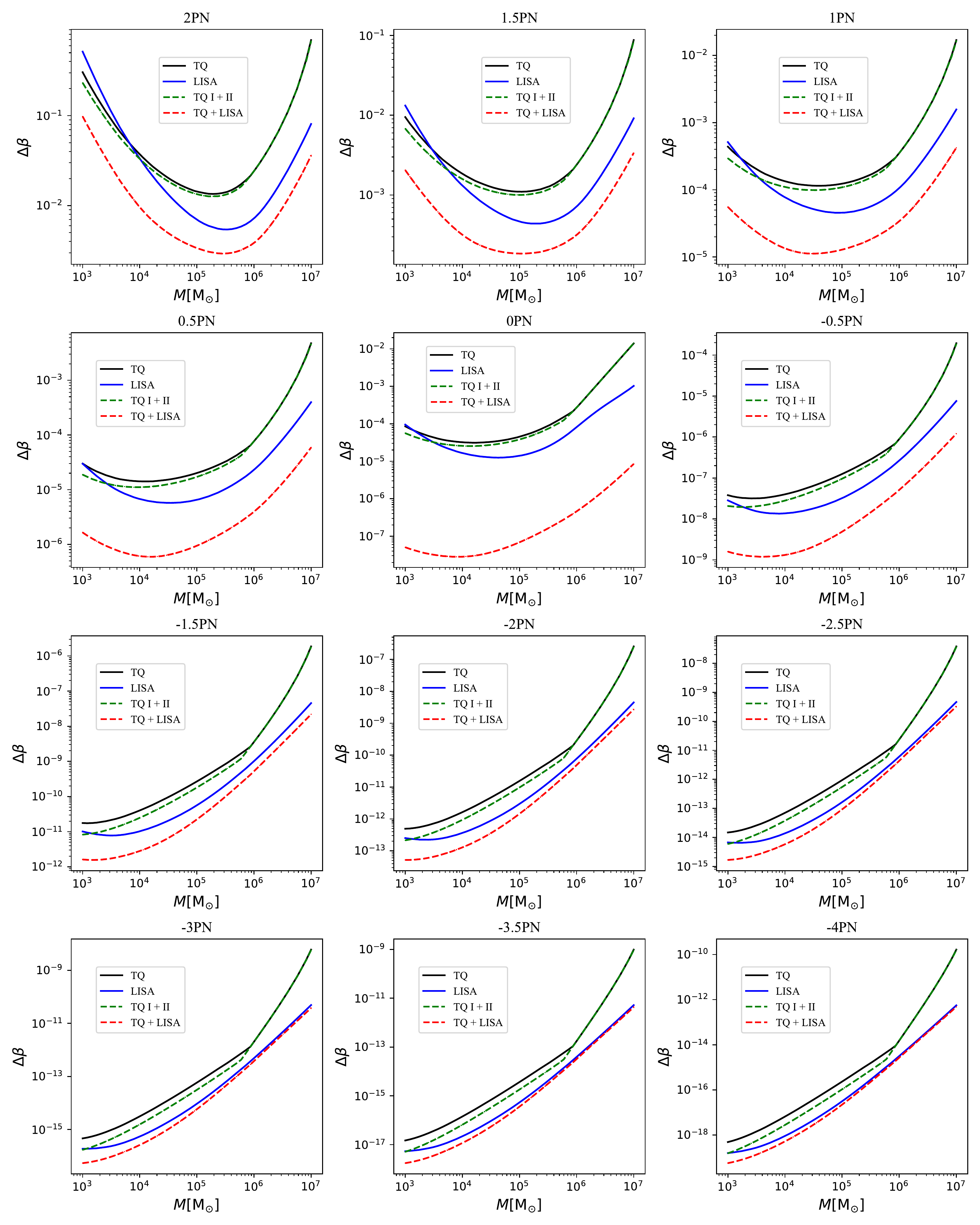}
	\caption{Dependence of $\Delta\beta$ on the total mass in the high mass range at different PN orders for different detector configurations.}
	\label{fig.spmassall}
\end{figure*}

\begin{figure*}
\centering
\includegraphics[width=1\textwidth]{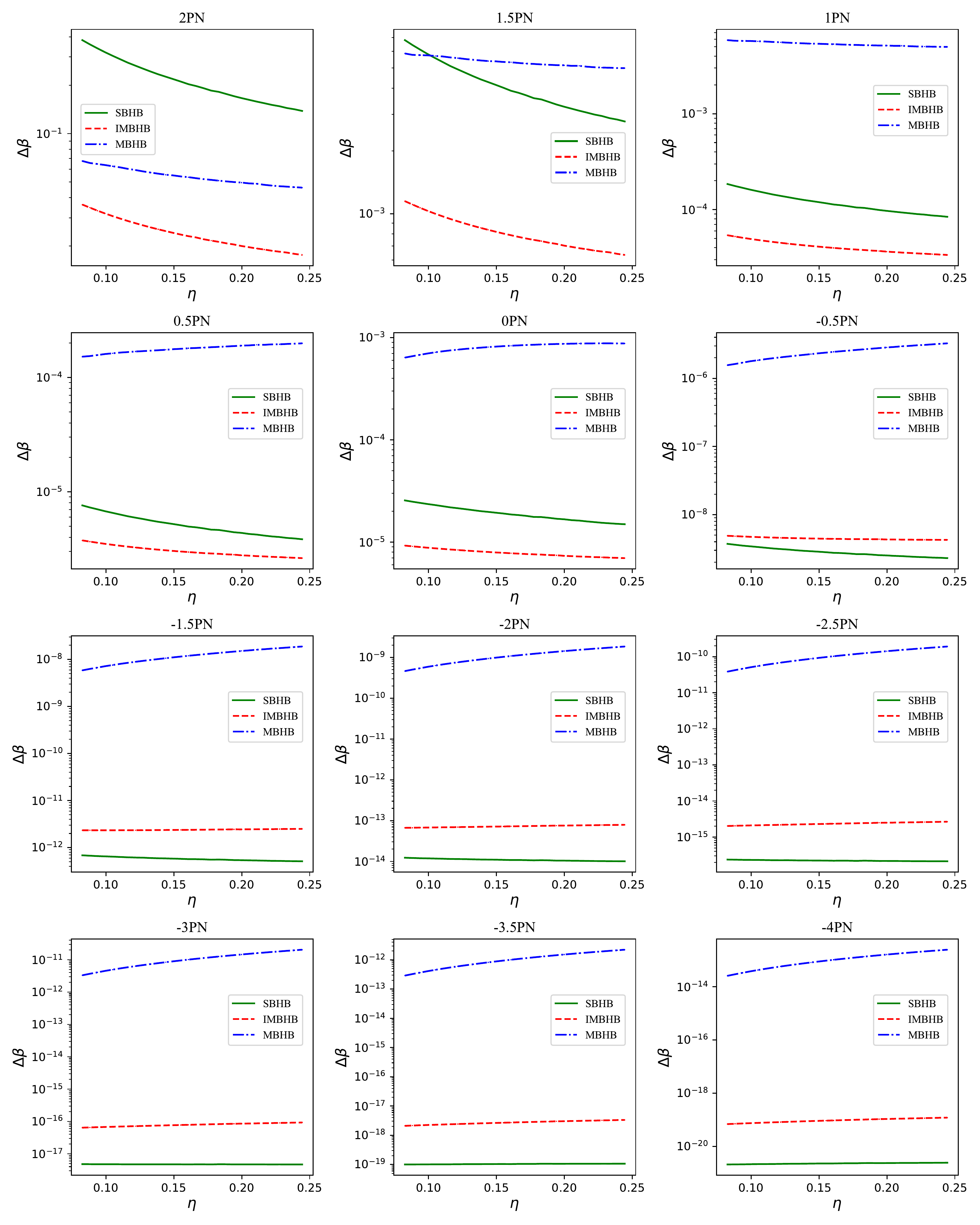}
\caption{Dependence of $\Delta\beta$ on the symmetric mass ration at different PN orders for TianQin.}
\label{fig.etadpall}
\end{figure*}

\begin{figure*}
\centering
\includegraphics[width=1\textwidth]{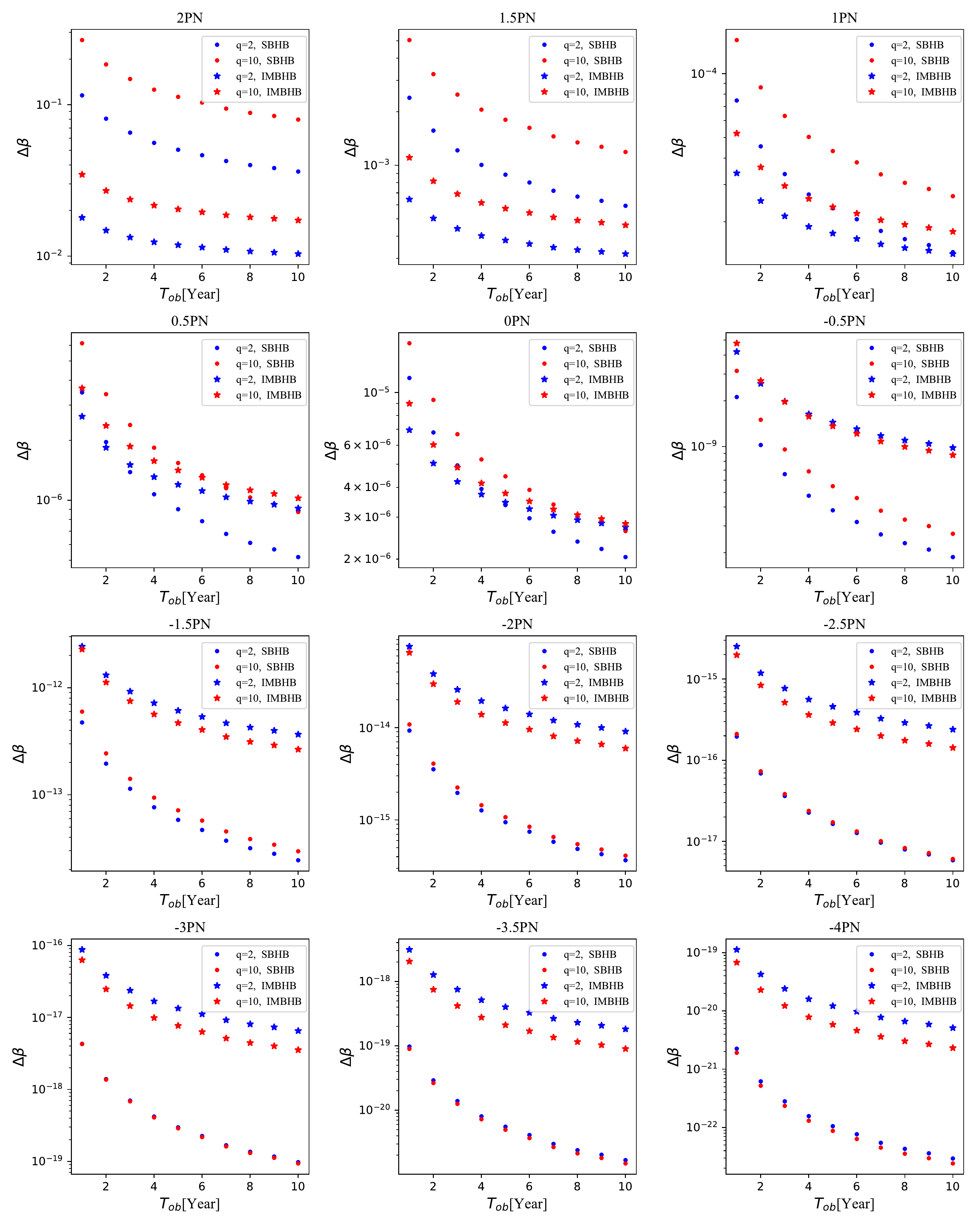}
\caption{Dependence of $\Delta\beta$ on the observation time at different PN orders for TianQin.}
\label{fig.tobdpall}
\end{figure*}

\begin{figure*}
\centering
\includegraphics[width=1\textwidth]{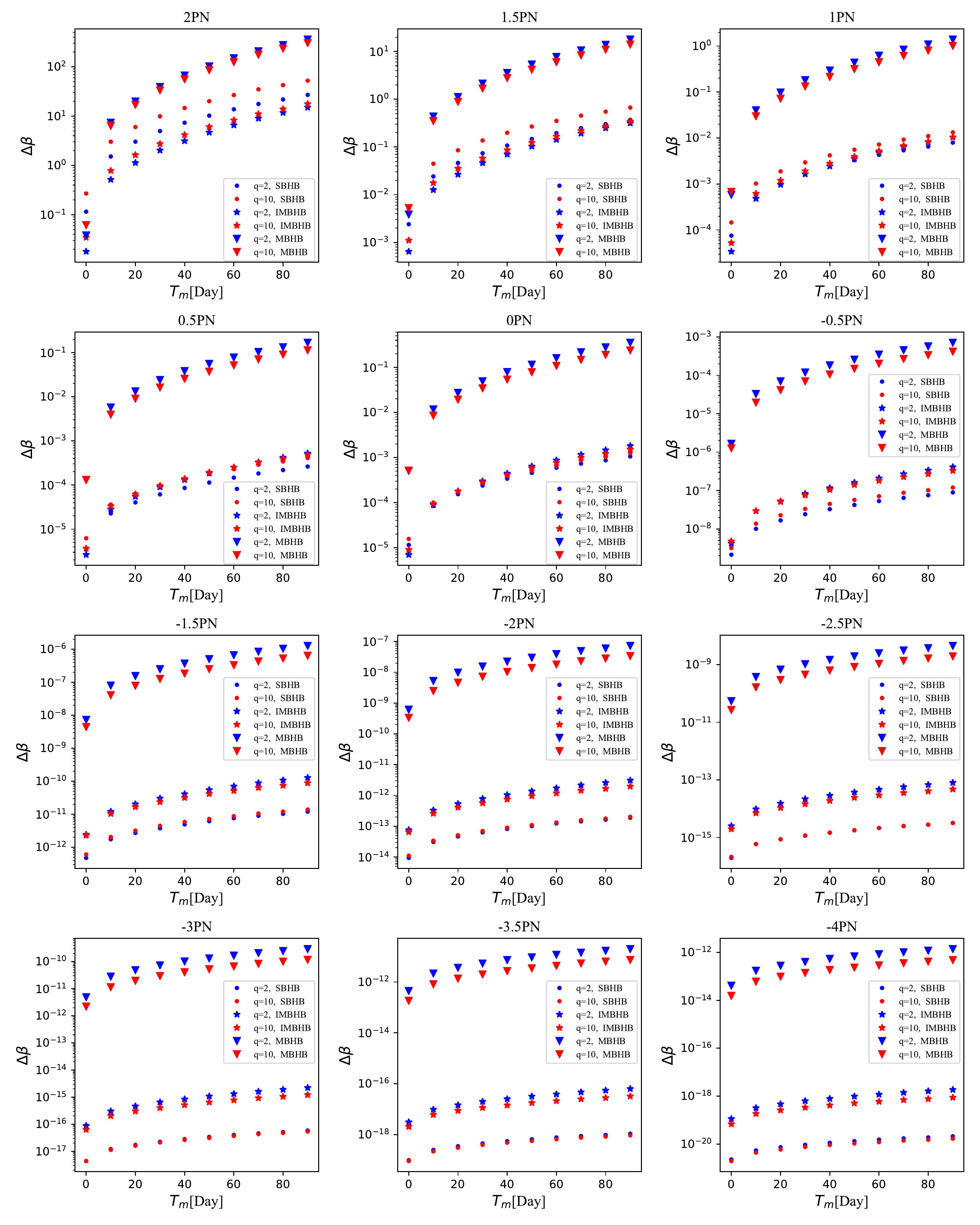}
\caption{Dependence of $\Delta\beta$ on the lost observation time at different PN orders for TianQin.}
\label{fig.tmdpall}
\end{figure*}

\begin{figure*}
\centering
\includegraphics[width=1\textwidth]{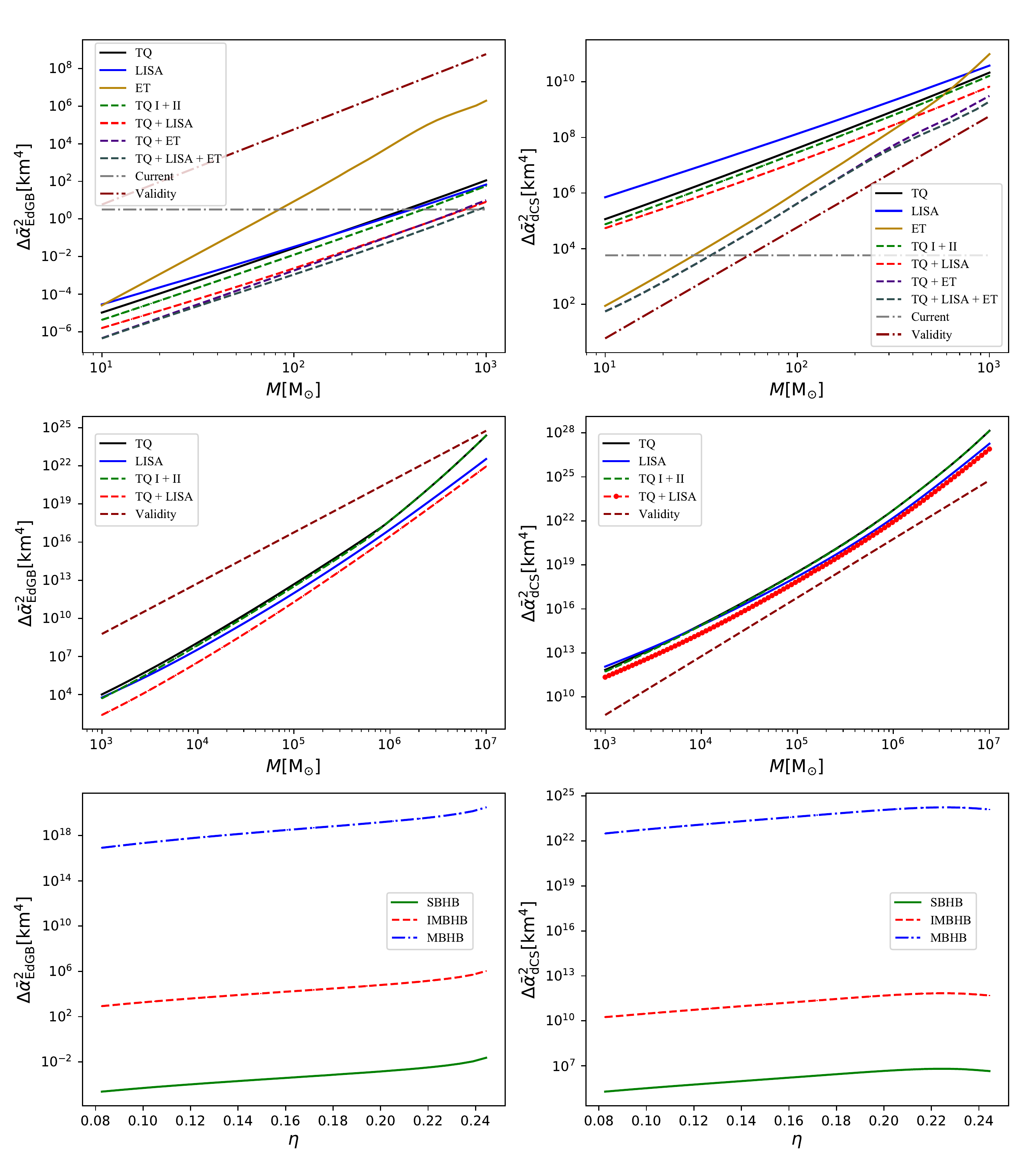}
\caption{Dependence of $\Delta\bar{\alpha}_{\rm EdGB}^2$ (left) and $\Delta\bar{\alpha}_{\rm dCS}^2$ (right) on the total mass and the symmetric mass ratio. ``Current" means the current best bound from GW detection. ``Validity" means the bound imposed by (\ref{validity}), and only results below the indicated bound are considered reliable.}
\label{fig.dcsedgb}
\end{figure*}

\end{center}
\end{widetext}
\end{document}